\documentclass[aps,10pt,superscriptaddress]{revtex4-2}
\usepackage{graphicx} %

\usepackage{physics}
\usepackage{amsfonts}
\usepackage{amsmath}
\usepackage{hyperref}
\usepackage{amssymb}
\usepackage{xcolor}
\usepackage{bbm}
\usepackage{cancel}
\usepackage{subfig}

\newcommand{\pt}[1]{\left( #1 \right)}
\newcommand{\pq}[1]{\left[ #1 \right]}
\newcommand{\pg}[1]{\left\{ #1 \right\}}
\newcommand{\pgt}[1]{\{ #1 \}}

\newcommand{\id}{\hat{\mathbbm{1}}}
\newcommand{\nohatid}{\mathbbm{1}}

\newcommand{\opxprod}[1]{{#1}^{\dag}{#1}}

\newcommand{\gikl}{\gamma^{\pt{l,i,k}}}
\newcommand{\giklp}{\gamma^{\pt{l,i,k'}}}

\newcommand{\apa}[2]{{a'}_{#1}^{*}a_{#2}}
\newcommand{\Xtilde}[1]{\Tilde{X}_{#1}}
\newcommand{\Ytilde}[1]{\Tilde{Y}_{#1}}
\newcommand{\Ztilde}[1]{\Tilde{Z}_{#1}}

\begin{document}

\preprint{APS/123-QED}

\title{Loss-tolerant quantum key distribution with detection efficiency mismatch}

\author{Alessandro Marcomini}
\email{amarcomini@vqcc.uvigo.es}
\affiliation{Vigo Quantum Communication Center, University of Vigo, Vigo E-36310, Spain}
\affiliation{Escuela de Ingeniería de Telecomunicación, Department of Signal Theory and Communications, University of Vigo, Vigo E-36310, Spain}
\affiliation{AtlanTTic Research Center, University of Vigo, E-36310, Spain}
\author{Akihiro Mizutani}
\affiliation{Faculty of Engineering, University of Toyama, Gofuku 3190, Toyama 930-8555, Japan}
\author{Fadri Gr\"unenfelder}
\affiliation{Vigo Quantum Communication Center, University of Vigo, Vigo E-36310, Spain}
\affiliation{Escuela de Ingeniería de Telecomunicación, Department of Signal Theory and Communications, University of Vigo, Vigo E-36310, Spain}
\affiliation{AtlanTTic Research Center, University of Vigo, E-36310, Spain}
\author{Marcos Curty}
\affiliation{Vigo Quantum Communication Center, University of Vigo, Vigo E-36310, Spain}
\affiliation{Escuela de Ingeniería de Telecomunicación, Department of Signal Theory and Communications, University of Vigo, Vigo E-36310, Spain}
\affiliation{AtlanTTic Research Center, University of Vigo, E-36310, Spain}
\author{Kiyoshi Tamaki}
\affiliation{Faculty of Engineering, University of Toyama, Gofuku 3190, Toyama 930-8555, Japan}

\date{\today}

\begin{abstract}
    {Current implementations of quantum key distribution (QKD) typically rely on prepare-and-measure (P\&M) schemes.
    Unfortunately, these implementations are not completely secure, unless security proofs fully incorporate all imperfections of real devices. So far,
    existing {proofs} have primarily focused on imperfections of either the light source or the measurement device.
    In this paper, we establish a security proof for the loss-tolerant P\&M QKD protocol that incorporates imperfections in both the source and the detectors. Specifically, we demonstrate the security of {this scheme} when the emitted states deviate from the ideal {ones} and Bob’s measurement device does not meet the basis-independent detection efficiency condition. Furthermore, we conduct an experiment to characterise the detection efficiency mismatch of commercial {single-photon} detectors as a function of the polarisation state of the input light, and determine the expected secret key rate in the presence of state preparation flaws when using such detectors.  
    Our work provides a way towards {guaranteeing} the security of actual implementations of widely deployed P\&M QKD.}
\end{abstract}

\maketitle
\section{\label{sec:Intro} Introduction}
Quantum key distribution (QKD) offers the potential for communication certified with information-theoretic security by utilizing the principles of quantum mechanics. Specifically, it enables two legitimate parties, commonly referred to as Alice and Bob, to generate a shared symmetric secret key, regardless of any eavesdropping effort by a third party (Eve) \cite{loSecureQuantum2014, xuSecureQuantum2020, pirandolaAdvancesQuantum2020}.

As of today, significant results have been achieved to show how QKD can be deployed both in metropolitan and inter-city networks \cite{dynesCambridgeQuantum2019, stuckiLongtermPerformance2011, sasakiFieldTest2011,chenIntegratedSpacetoground2021} and over ground-to-satellite links \cite{liaoSatellitetogroundQuantum2017, chenIntegratedSpacetoground2021}, as well as in commercial systems \cite{HttpsWwwb, HttpsWww, HttpsWwwa}. Still, the widespread of this technology is challenged by the fact that real-world devices present imperfections, potentially implying unnoticed security loopholes or information leakages. 
This fact is remarked by the richness of literature in the field of quantum hacking \cite{BSI, jainAttacksPractical2016, sunReviewSecurity2022}, by examples of successful attacks \cite{zhaoQuantumHacking2008, lydersenHackingCommercial2010, gerhardtFullfieldImplementation2011, weierQuantumEavesdropping2011, jainDeviceCalibration2011} and continuously proposed malicious schemes \cite{pangHackingQuantum2020, tanExternalMagnetic2022, yeInducedPhotorefractionAttack2023, wuHackingSinglephoton2020, zhangAnalysisInjectionlockingloophole2022, baliukaDeeplearningbasedRadiofrequency2023}. As a consequence, introducing security proofs for QKD with realistic devices has become of paramount importance \cite{zapateroImplementationSecurity}. 

The most vulnerable {part} of a QKD setup is widely acknowledged to be the receiver. This is mainly due to two reasons. Firstly, the receiver is, by definition, welcoming signals from the quantum channel. As a consequence, Eve can send additional light to the detectors, inducing information leakage (e.g., through the so-called Trojan-horse attack \cite{sajeedInvisibleTrojanhorse2017, jainTrojanhorseAttacks2014}), forcing a convenient operational regime of the detectors (e.g., via an after-gate attack for gated detectors \cite{wiechersGateAttack2011}) or directly blinding them through bright light \cite{lydersenHackingCommercial2010, gerhardtFullfieldImplementation2011}. Secondly, even if the aforementioned attacks could be prevented \cite{yuanResilienceGated2011,yuanResponseResilience2011}, the response of single-photon detectors (SPDs) typically depends on several degrees of freedom, such as time of arrival, frequency, polarisation and spatial modes of the incoming pulses \cite{sajeedSecurityLoophole2015, rauSpatialMode2015, chaiwongkhotEavesdroppersAbility2019, weiImplementationSecurity2019}. This allows Eve to manipulate the detection probability of the SPDs by changing the properties of the {arriving} signals \cite{zhaoQuantumHacking2008, jainDeviceCalibration2011}. 
Moreover, as QKD protocols require to discriminate different states at the receiver, multiple SPDs are {usually} {needed. Since} realistic devices cannot be identical, Eve might modify the incoming quantum states in such a way that a specific outcome detection is highly favoured. {The latter issue} is commonly addressed as detection efficiency mismatch (DEM) \cite{fungSecurityProof2009, lydersenSecurityOf2008, maOperationalInterpretation2019, bochkovSecurityOf2019, zhangSecurityProof2021, trushechkinSecurityOf2022}. 

On the transmitter side, security threats are typically less concerning. For example, issues related to multi-photon emissions can be mitigated by means of the decoy-state method \cite{hwangQuantumKey2003, wangBeatingPhotonNumberSplitting2005, loDecoyState2005, maPracticalDecoy2005, limConciseSecurity2014}, while the effect of state preparation flaws (SPFs) has been shown to be almost negligible thanks to the so-called loss-tolerant (LT) approach \cite{tamakiLosstolerantQuantum2014} (although the latter crucially requires identical detectors). The effect of side-channels has also been intensively studied \cite{pereiraQuantumFlawedLeaky2019,pereiraQuantumKey2020, zapateroSecurityQuantum2021, curras-lorenzoSecurityQuantum2024, curras-lorenzoSecurityFramework2023}. 

As a result, in the last decade huge effort has been dedicated to the design of QKD protocols based on quantum interference that could lift all assumptions on the receiver side, namely measurement-device-independent (MDI) \cite{loMeasurementDeviceIndependentQuantum2012} and twin-field (TF) QKD \cite{lucamariniOvercomingRate2018}.
While on-field realisations of these schemes have been achieved over large distances \cite{wangTwinfieldQuantum2022, liuExperimentalTwinField2023, liuExperimentalDemonstration2023}, prepare-and-measure (P\&M) protocols are still being preferred for short- and mid-range communication and for industrial products due to their simpler implementation. Thus, the development of security proofs that can jointly account for both transmitter and receiver imperfections is of {great relevance}.

In this work we extend the security proof for LT QKD to accommodate {the existence of a} detection efficiency mismatch. In doing so, we provide a unique recipe to establish the asymptotic security of P\&M QKD protocols against coherent attacks in the presence of both SPFs and DEM. Moreover, we conduct an experiment to characterise {the DEM in commercial {SPDs} and show that a secret key can be generated using such devices, even in the presence of large SPFs.}
In detail, we adopt the complementarity framework introduced by Koashi \cite{koashiSimpleSecurity2009} to provide an upper bound on the phase-error rate, which ultimately quantifies the amount of privacy amplification {that} Alice and Bob must perform in the postprocessing {phase of the QKD protocol}. Our idea generalises the LT approach to the case in which {the detection probability associated to the two measurement bases differs}. In particular, we show that characterising the efficiency of the SPDs and monitoring the send-and-receive statistics of the protocol is enough to directly estimate all quantities involved in the computation of the secret key rate (SKR). In so doing, we generalise and simplify the technique developed in Ref.~\cite{fungSecurityProof2009}.

The paper is organised as follows. In Sec.~\ref{sec: assumptions} we illustrate the underlying assumptions for our analysis, while in Sec.~\ref{sec: protocol} we present the details of the {QKD scheme} we evaluate. In Sec.~\ref{sec: main theorem} we report the main results of our study and provide a full procedure for the estimation of the SKR. We show the applicability of our analysis by performing an experimental characterisation of two real detectors and studying how the DEM affects performance in Sec.~\ref{sec: experiment}. Finally, we discuss the implications of our work in Sec.~\ref{sec: discussion} and conclude with some remarks in Sec.~\ref{sec: conclusions}. 
Further calculations and discussions can be found in the Appendices.

\section{Assumptions}\label{sec: assumptions}
Our analysis relies on the following assumptions, whose implications are further discussed in Appendix~\ref{sec: setup model}.
\begin{itemize}
    \item[(A1)] Alice has a perfect single-photon source and she knows the qubit states $\pg{\ket{\phi_i}_B}_i$, with $i\in\pg{0_Z,1_Z,0_X}$, that she is sending to Bob. These might be different from the perfect $\ket{i}_B$ states (i.e., the eigenstates of the $\hat{Z}$ and $\hat{X}$ Pauli matrices). Moreover, we assume the flawed states $\pg{\ket{\phi_i}_B}_i$ to be pairwise linearly independent and to lay on the $XZ-$plane of the Bloch sphere (which is always true, up to a lifting filtering operation, as discussed in Appendix~\ref{sec: receiver}).
    \item[(A2)] {Alice's lab is perfectly shielded from the eavesdropper such that no unwanted information leakage occurs (i.e., there is no side channel in Alice's source)}.
    \item[(A3)] Bob employs an active BB84 receiver with two detectors. This means that the detectors are placed after a basis selector such that the detector $D_0$ ($D_1$) measuring $0_Z$ ($1_Z$) is also employed to measure $0_X$ ($1_X$). 
    \item[(A4)] Bob receives the system $B$ in a qubit space.
    \item[(A5)] {As illustrated in Appendix~\ref{sec: receiver}, Bob's {positive operator-valued measure (POVM) element} assigning outcome $s\in\pg{0,1}$ in the basis $\beta\in\pg{X,Z}$ has the form}
    \begin{equation}\label{eqn: full BT POVM definitions in assumptions}
        \hat{M}^{s_{\beta}}_{BT} = \dyad{s_{\beta}}_{B} \otimes \pt{\opxprod{\hat{F}_s}}_T,
    \end{equation} 
    {while the POVM element associated to a failed detection in the basis $\beta$ is given by 
    \begin{equation}\label{eqn: fail POVM in assumptions}
        \hat{M}^{{fail}_{\beta}}_{BT} = \id - \hat{M}^{0_{\beta}}_{BT} - \hat{M}^{1_{\beta}}_{BT}.
    \end{equation}}
    Here, $T$ denotes an additional system of arbitrary dimension corresponding to a given efficiency-affecting mode (e.g., {time of arrival, frequency, polarisation or spatial mode}), while $\opxprod{\hat{F}_s}$ denotes the POVM element for such system corresponding to the detector $D_s$ clicking \cite{fungSecurityProof2009}. In this respect, $\opxprod{\hat{F}_s}$ takes the role of a generalised efficiency operator. {Note that when $\opxprod{\hat{F}_s} = \eta_s\id$, with $\eta_s \in [0,1]$, we obtain} a model for detectors with constant efficiencies, which might be different for $D_0$ and $D_1$. While this {description} might not be the most general one, it well captures how the detectors functioning depends on additional modes and allows to investigate any scenario in which the information-carrying degree of freedom can be decoupled from the efficiency-affecting one.
    
    We assume Bob can characterise his two detectors by performing an efficiency tomography {measurement, and thus he learns} the POVM elements $\{\opxprod{\hat{F}_s}\}_{s}$. We also assume $\hat{F}_0$ and $\hat{F}_1$ to be invertible, as no positive {SKR} can be achieved when this requirement is not satisfied \cite{fungSecurityProof2009}.
    \item[(A6)] We consider the asymptotic case where Alice sends to Bob an infinite number of signals and investigate coherent attacks. 
\end{itemize}

Assumptions (A1)-(A2) imply that Alice knows precisely the states that are sent to the channel in each round of the protocol. On the receiver side, Assumptions (A3)-(A4)-(A5) imply that Bob's POVMs are known and are defined over the tensor product space of systems $B$ and $T$, being the former a qubit space.
Importantly, we remark that our work is fully compatible with the decoy-state method for QKD \cite{hwangQuantumKey2003, wangBeatingPhotonNumberSplitting2005, loDecoyState2005, maPracticalDecoy2005, limConciseSecurity2014}, as the latter allows to estimate relevant probabilities related to single-photon signals.

\section{Protocol description}\label{sec: protocol}

In this section we introduce the fundamentals of the P\&M three-state protocol which we base our analysis upon. We adopt the LT {scheme, which considers basis mismatched events,} and shall refer to it as the {\it actual protocol} in the following. Note that in the asymptotic regime, the phase-error rate for this scheme is the same as {for} the standard BB84 protocol \cite{tamakiLosstolerantQuantum2014}. 

Together with the actual protocol, we introduce a \textit{virtual protocol} which is completely equivalent from Eve's viewpoint and allows for the phase-error rate estimation. Importantly, we fictitiously consider this protocol only for the purpose of the security proof and there is no need to actually implement it.

\vspace{10pt}
{\it Actual Protocol}
       \begin{enumerate}	
	    \item State preparation: Alice selects the $Z$ ($X$) basis with probability $p_{Z_A}$ ($p_{X_A}$). When the $Z$ basis is selected, she prepares a single-photon pulse in the state $\ket{\phi_{0_Z}}_B$ or $\ket{\phi_{1_Z}}_B$ uniformly at random, whereas a single-photon pulse is prepared in the state $\ket{\phi_{0_X}}_B$ for the $X$ basis selection. Then, she sends the single-photon pulse to Bob through a quantum channel. She repeats this step $N$ times. 
     
     Together with the $B$ system, Bob receives from the channel an auxiliary $T$ system which encodes information about the efficiency-affecting mode.
	             
	    \item Measurement: For each pulse he receives, Bob selects the $Z$ ($X$) basis with probability $p_{Z_B}$ ($p_{X_B}$).  Then, when the $Z$ ($X$) basis is selected, he performs a measurement $\hat{M}_{BT}^{Z}$ ($\hat{M}_{BT}^{X}$) described by the {POVM} $\{{\hat M}^{0_Z}_{BT}, {\hat M}^{1_Z}_{BT},$ ${\hat M}^{fail_Z}_{BT}\}$ $\pt{\pgt{{\hat M}^{0_X}_{BT}, {\hat M}^{1_X}_{BT}, {\hat M}^{fail_X}_{BT}}}$ over systems $B$ and $T$. The POVM elements ${\hat M}^{s_{\beta}}_{BT}$ for $s\in\pg{0,1}$ and $\beta\in\pg{X,Z}$ are described by Eq.~\eqref{eqn: full BT POVM definitions in assumptions}. The measurement outcomes corresponding to these operators are defined as detection events. On the other hand, ${\hat M}^{fail_{\beta}}_{BT}$ corresponds to an inconclusive event in the basis $\beta$. Note that due to DEM we have that ${\hat M}^{fail_{X}}_{BT}\neq{\hat M}^{fail_{Z}}_{BT}$.
	    
	    \item Sifting: Once the previous steps are completed, Bob announces to Alice over an authenticated public channel in which rounds he obtained detection events, as well as the basis selection he made for each of these. This way, they can compute the sifted key, defined as the bit string generated from the instances in which Alice selected the $Z$ basis and Bob obtained a detection event in the $Z$ basis. 
        {Bob also shares with Alice a fraction of his results, allowing her to compute the {probabilities} associated to each possible detection event for each type of signal sent. Formally, we denote {as $p_{s_\beta, i}$ the joint probability of Bob obtaining an outcome $s_{\beta}$, with $s\in\pgt{0,1}$ and $\beta\in\pgt{X,Z}$, and Alice sending the signal $i\in\pgt{0_Z,1_Z,0_X}$}.
     
	    \item Parameter estimation: By using the {estimated probabilities for measurements in the $X$ basis $\{p_{s_X, i}\}_{s,i}$}, Alice estimates an upper bound on the phase-error rate of the sifted key, while {she estimates the bit-error rate in the $Z$ basis through the values of $\{p_{s_Z, i}\}_{s,i}$ for $i\in\pg{0_Z,1_Z}$}. These are the crucial parameters for privacy amplification and error correction, respectively. }
     
	    \item Data-postprocessing: Through discussions over an authenticated public channel, Alice and Bob perform error correction, and then they conduct privacy amplification to generate a secret key.	
	\end{enumerate}
 
     The following virtual scheme is based on an equivalent entanglement-based (or source-replacement) approach. Crucially, Eve cannot distinguish this protocol from the actual one, as both the quantum and classical information available to Eve is the same for both protocols. Moreover, all the data that Alice and Bob use for the phase-error rate estimation (that is, all of Alice and Bob's detection data except for the sifted key) are also the same for both schemes. As a result, the phase-error rate estimated from the virtual protocol can be employed to determine the amount of privacy amplification needed to generate a {secret} key in the actual protocol \cite{koashiSimpleSecurity2009, loSecurityQuantum2007}.

\newpage
{\it Virtual Protocol}
       \begin{enumerate}	
	    \item Distribution of entanglement: Alice prepares bipartite systems, each of which is composed of a shield system $A$ and a single-photon pulse $B$, in a joint entangled state in the form
        \begin{equation}\label{eqn:     varphi_AB}
        \ket{\varphi}_{AB} = \sum_{c=1}^6 \sqrt{p_c} \ket{c}_{A}\ket{\phi^{\pt{c}}}_B.
        \end{equation}

        Each projection of the shield system onto one of the orthonormal basis states $\{\ket{c}_{A}\}_{c\in\pg{1,\ldots,6}}$ corresponds to a different configuration of the state $\ket{\phi^{\pt{c}}}_B$ sent to Bob and of the measurement basis choice for him. Specifically,
        Alice's measurement outcome being $c\in\{1,2\}$ corresponds to the rounds in which both her and Bob select the $Z$ basis. For the sake of estimating the phase-error rate, in these rounds Bob is allowed to replace his standard $Z$ basis measurement with a so-called virtual $X$ basis measurement over systems $B$ and $T$ \cite{koashiSimpleSecurity2009}. Details on the analytical form of this measurement are reported below.
        When Alice's outcome is $c\in\{3, 4, 5\}$, she is actually sending the states $\ket{\phi_{0_Z}}_B$, $\ket{\phi_{1_Z}}_B$ and $\ket{\phi_{0_X}}_B$, respectively, while Bob performs the same $X$ basis measurement $\hat{M}^X_{BT}$ {as in} the actual protocol. Finally, Alice's outcome $c=6$ corresponds to her sending $\ket{\phi_{0_X}}_B$ and Bob measuring in the $Z$ basis.
        Further details on this state preparation {procedure} can be found in Appendix~\ref{sec: transmitter} and Table~\ref{tab: virtual protocol recap}. 

        {After the preparation, Alice sends the system $B$ to Bob through a quantum channel, while keeping the shield system in her lab. She
        repeats this step $N$ times.}
        
	    \item Delayed state preparation and measurement: After transmission, Alice performs the projective measurement described by $\{\dyad{c}_{A}\}_{c}$ over each of the $N$ systems $A$ she holds. 
        \item Announcement of detection events and bases for sifting, parameter estimation and postprocessing: These phases are tackled as in the actual protocol. One can follow the scheme {provided} in Fig.~\ref{fig:fig_scheme_c_states} to relate {the quantities $\{p_{s_{\beta},i}\}_{s,\beta,i}$} introduced in the actual protocol to the statics of the virtual protocol for various values of $c$. 
	\end{enumerate}

Bob's virtual $X$ basis measurement {includes} two filtering operations acting on systems $B$ and $T$, followed by a projective measurement on system $B$. We further elaborate on filtering operations in Sec.~\ref{sec: filtering}. In detail, we consider Bob first performing the filtering described by $\{ {\hat Q}_Z= \sqrt{\id-{\hat M}^{fail_Z}_{BT}}, \sqrt{{\hat M}^{fail_Z}_{BT}} \}$, followed by another (virtual) filtering operation $\{{\hat G}_{BT}, \sqrt{{\id}-{\hat G}_{BT}^{\dagger}{\hat G}_{BT}} \}$ and finally by the {projective} measurement $\{\dyad{0_X}_B\otimes {\id}_T, \dyad{1_X}_B\otimes {\id}_T\}$. The analytical form of ${\hat G}_{BT}$ is introduced in Eq.~\eqref{eqn: virtual filter introduction}. 

\subsection{Filtering operations}\label{sec: filtering}
Formally, a filtering operation is defined as a set of two Kraus operators $\pg{\hat{Q}_{s},\hat{Q}_{f}}$ (respectively, success and fail operators) acting on a given system $S_1$ in a state $\ket{x}_{S_1}$ such that, with a certain probability $p_s\le 1$, the final state of the system is given by $\hat{Q}_{s}\ket{x}_{S_1}$ {\cite{BennettMixedstateEntanglement1996}}. This is equivalent to considering an additional system $S_2$ and defining an unitary operation on the joint system $S_1\otimes S_2$ in the form
\begin{equation}\label{eqn: filtering definition}
    U\ket{x}_{S_1}\ket{0}_{S_2}=\hat{Q}_s \ket{x}_{S_1}\otimes\ket{s}_{S_2}+\hat{Q}_f \ket{x}_{S_1}\otimes\ket{f}_{S_2},
\end{equation}
for any state $\ket{x}_{S_1}$. Here the (arbitrary) orthonormal states $\ket{s}_{S_2}$ and $\ket{f}_{S_2}$ for system $S_2$ denote success and failure of the filter, respectively. Crucially, this notion highlights how measuring the system $S_2$ after applying the filtering allows to immediately assert whether the filtering was successful. 
It follows from Eq.~\eqref{eqn: filtering definition} that the probability of successful filtering can be found as
\begin{equation}
    p_s := \expval{\opxprod{\hat{Q}_s}}{x}_{S_1}.
\end{equation}

Importantly, through this paper we will address as ``filter" to a success operator $\hat{Q}_{s}$, therefore specifying only the evolution of a system in the case of successful filtering, since failure events are discarded. Note that for a filter to be valid it is sufficient that $\hat{Q}_{s}\hat{Q}_{s}^{\dag} \le \id$ (equivalently, $\hat{Q}_{s}^{\dag}\hat{Q}_{s} \le \id$), as this implies the existence of a valid corresponding failure operator and thus guarantees the non-violation of the unitarity of quantum mechanics.

Following the approach in Ref.~\cite{fungSecurityProof2009}, in this work we adopt the so-called Procrustean method of filtering \cite{bennettConcentratingPartial1996}. This choice is motivated by the fact that, due to {the DEM}, Bob's detectors do not project onto perfectly orthogonal states of the joint system $BT$. Procrustean filtering allows, with a success probability smaller than unity, to orthogonalise the initial states and therefore offers a convenient way to apply Koashi's security proof \cite{koashiSimpleSecurity2009}. More details on this matter can be found in Appendix~\ref{sec: receiver}.

\section{Main result}\label{sec: main theorem}
In this section we summarise the main results of our analysis, providing a straightforward way to compute the SKR from experimental data. The proof of these results can be found in Appendix~\ref{sec: result proof}. 

Let us consider the protocol introduced in Sec.~\ref{sec: protocol} and assume that the requirements in Sec.~\ref{sec: assumptions} are met. Let $\pmb{V}_i=\pt{V_{i}^{x},0,V_{i}^{z}}^T$ denote the Bloch vector of the state $\ket{\phi_i}_B$, with $i\in\pg{0_Z,1_Z,0_X}$. We define 
\begin{equation}
    p_3 = p_4 := \frac{p_{Z_A}p_{X_B}}{2}, \quad  \quad 
    p_5 := p_{X_A}p_{X_B},
\end{equation}
{and
\begin{equation}\label{eqn: main theorem LT q tilda}
    \pmb{\Tilde{q}}_s=
    \mqty(
    \Tilde{q}_{s_X, \nohatid} \\
    \Tilde{q}_{s_X, X} \\
    \Tilde{q}_{s_X, Z} \\
    )
    := 
    \mqty(
    p_3 &p_3V_{0_Z}^{x}&p_3V_{0_Z}^{z} \\
    p_4 &p_4V_{1_Z}^{x}&p_4V_{1_Z}^{z} \\
    p_5 &p_5V_{0_X}^{x}&p_5V_{0_X}^{z} \\
    )^{-1} 
    \mqty(
    p_{s_X,0_Z} \\
    p_{s_X,1_Z} \\
    p_{s_X,0_X}
    ),
\end{equation}}
for $s\in\pg{0,1}$. 

Now, consider the unitary diagonalisation of the following Gram matrix
\begin{equation}\label{eqn: main theorem diagonalization of FFFF}
\hat{F}_0\pt{\opxprod{\hat{F}_1}}^{-1}\hat{F}_0^{\dag} = \hat{U}\hat{D}\hat{U}^{\dag},
\end{equation}
such that $\hat{D} = diag\pt{D_1,D_2,\ldots}$, with $D_i \ge 0$ $\forall i$. Let us define 
\begin{equation}\label{eqn: C intro main thm}
    \hat{C} := \hat{C}_1 \hat{U}^{\dag} \hat{F}_0,
\end{equation}
where
\begin{equation}\label{eqn: def C1}
    \hat{C}_1 = diag\pt{\sqrt{\eta_1},\sqrt{\eta_2},\ldots} , \quad \quad \eta_i := \min\pt{\frac{1}{D_i}, 1}.
\end{equation}
Moreover, consider a set of four real, non-negative values
$\pmb{\Lambda}=\pg{\lambda^{-}_s,\lambda^{+}_s}_{s\in\pg{0,1}}$ such that
\begin{equation}\label{eqn: summary op bound for CdagC}
    \lambda^{-}_s \opxprod{\hat{F}_s} \le
     \opxprod{\hat{C}} \le 
    \lambda^{+}_s  \opxprod{\hat{F}_s},
\end{equation}
for $s\in\pg{0,1}$.

Then, for the general case of coherent attacks, the asymptotic secret key rate per channel use $R$ is lower bounded {as
\begin{equation}\label{eqn: main theorem SKR}
    R \ge 
    p_{Z}^{sift}\pq{r_X^{virt, L}\pt{1-h_2\pt{e_p^{{U}}}} - fh_2\pt{e_b}},
\end{equation}
where
\begin{equation}\label{eqn: N sift main thm}
    p_{Z}^{sift} := \sum_{s=0}^1\pt{p_{s_Z,0_Z}+p_{s_Z,1_Z}}
\end{equation}
is the probability of being in a sifted key round, the term $r_X^{virt, L}$ denotes} a lower bound on the fraction of sifted key rounds in which Bob's {virtual $X$ basis measurement} is successful (that is, the fraction of sifted key rounds for which Bob succeeds in making a guess for Alice's virtual $X$ measurement outcome), 
{$h_2\pt{x} = -x\log_2(x) - (1-x)\log_2(1-x)$ denotes the binary entropy, $e_p^{{U}}$ is an upper bound on the phase-error rate, $f$ is the efficiency parameter associated to the error correction process and} $e_b$ is the bit-error rate. 
{The latter quantity can be directly inferred from the 
estimates of the detection probabilities in the sifted key rounds and is given by
\begin{equation}\label{eqn: e_b main thm}
    e_b := \frac{p_{0_Z,1_Z} + p_{1_Z,0_Z}}{p_Z^{sift}}.
\end{equation}}
As for the bounds $r_X^{virt, L}$ and $e_p^{{U}}$, let us first {define 
\begin{equation}\label{eqn: main theorem N_x_virt}
    p_{X}^{virt,L} := p_{Z_A}p_{Z_B}\pq{\braket{\phi_{0_Z}}{\phi_{1_Z}}\pt{\frac{\lambda_0^+ +\lambda_0^-}{2}\Tilde{q}_{0_X,X} + \frac{\lambda_1^+ +\lambda_1^-}{2}\Tilde{q}_{1_X,X}}
    -\pt{\frac{\lambda_0^+ - 3\lambda_0^-}{2} \Tilde{q}_{0_X,\nohatid} +\frac{\lambda_1^+ - 3\lambda_1^-}{2}\Tilde{q}_{1_X,\nohatid}}},
\end{equation}
corresponding {to a lower bound on the average of the conditional probabilities of the filters in Bob's virtual $X$ measurement introduced in Sec.~\ref{sec: protocol} being successful, given Alice and Bob's outcomes in the previous rounds of the virtual protocol.}  
Then, we have {that
\begin{equation} 
    r_X^{virt, L} =
    \frac{p_{X}^{virt,L}}{p_Z^{sift}},
    \label{eqn: main thm bound p_s}
\end{equation}
while for {the upper bound on the phase-error rate we have that
\begin{align}\nonumber
    e_p^{{U}} = \frac{p_{X^{virt}}^{err, U}}{p_{X}^{virt,L}} &:= \frac{1}{p_{X}^{virt,L}}\left[\frac{1-\braket{\phi_{0_Z}}{\phi_{1_Z}}}{2}\pt{\frac{3\lambda_0^+ - \lambda_0^-}{2}\Tilde{q}_{0_X,\nohatid} - \frac{\lambda_0^+ + \lambda_0^-}{2}   \Tilde{q}_{0_X,X}} \right.
\nonumber \\ 
    & \ \ \quad \quad \quad + \left. \frac{1+\braket{\phi_{0_Z}}{\phi_{1_Z}}}{2}\pt{\frac{3\lambda_1^+ - \lambda_1^-}{2}\Tilde{q}_{1_X,\nohatid} + \frac{\lambda_1^+ + \lambda_1^-}{2}\Tilde{q}_{1_X,X}}\right] \label{eqn: main thm bound e_p},
\end{align}
where $p_{X^{virt}}^{err, U}$ denotes an upper bound on the average of the conditional probabilities of Bob making an error in his predictions in the
virtual $X$ measurement.

Although any choice of $\pmb{\Lambda}$ satisfying Eq.~\eqref{eqn: summary op bound for CdagC} is a valid choice, one should pick the values that maximise the SKR. Nevertheless, as Eqs.~\eqref{eqn: main theorem N_x_virt}-\eqref{eqn: main thm bound e_p} depend on the experimental {data through the estimates of $\{p_{s_{\beta},i}\}_{s,\beta,i}$ (and, thus,  $\pmb{\Tilde{q}}_s$}), finding the optimal set $\pmb{\Lambda}$ might be non-trivial. {Nevertheless, we note that the term $p_X^{virt, L}$} in Eq.~\eqref{eqn: main theorem N_x_virt} appears both at the numerator of Eq.~\eqref{eqn: main thm bound p_s} and at the denominator of Eq.~\eqref{eqn: main thm bound e_p}, and thus we conjecture that maximising this quantity over the possible values of $\pmb{\Lambda}$ corresponds to an optimisation of the SKR. This has a great impact since {$p_X^{virt, L}$}
is a linear combination of the parameters in $\pmb{\Lambda}$ and, crucially, the constraints on $\pmb{\Lambda}$ given by Eq.~\eqref{eqn: summary op bound for CdagC} are semidefinite constraints on linear operators. Therefore, one can find suitable values of $\pmb{\Lambda}$ by employing semidefinite programming, which allows for a fast and efficient optimisation. In detail, the semidefinite program (SDP) to solve is defined as
\begin{equation}\label{eqn: SDP main thm}
    \begin{array}{rl}
        \text{maximise} & p_X^{virt, L}, \\
        &\\
        \text{subject to} & \lambda^-_s \opxprod{\hat{F}_s} \le \opxprod{\hat{C}}, \ s\in\pg{0,1}, \\
        &   \opxprod{\hat{C}} \le \lambda^+_s  
        \opxprod{\hat{F}_s}, \ s\in\pg{0,1}, \\
        &   p_{X}^{virt,L} \le p_Z^{sift}, \\
        &   p_{X^{virt}}^{err, U} \le \tfrac{1}{2}p_{X}^{virt,L}. 
    \end{array}
\end{equation}
The second to last condition of the SDP above implies that $r_X^{virt, L}\le1$, which is due to the fact that Bob performs a virtual $X$ measurement for the sifted key rounds and therefore the probability of being in these rounds upper bounds the probability of Bob's virtual $X$ measurement being successful.
The last condition implies that the maximum phase-error rate allowed is {$1/2$}, corresponding to a uniformly random guess.

As an alternative to solving the SDP {in Eq.~\eqref{eqn: SDP main thm}}, in Appendix~\ref{sec: result proof} we prove that valid lower and upper analytical bounds for $\opxprod{\hat{C}}$ in Eq.~\eqref{eqn: summary op bound for CdagC}, although possibly suboptimal, are obtained for
\begin{gather}
\lambda^{-}_0 = \min\pt{D_{max}^{-1},1} , \quad
\lambda^{+}_0 = \min\pt{D_{min}^{-1},1} , \nonumber \\
\lambda^{-}_1 = \min\pt{D_{max}^{-1},1}D_{min}                 , \quad 
\lambda^{+}_1 =  \min\pt{D_{min}^{-1},1}D_{max},             
\label{eqn: AB bounds for lambda main thm}
\end{gather}
where $D_{max}$ ($D_{min}$) is the maximum (minimum) eigenvalue of the operator $\hat{D}$ introduced in Eq.~\eqref{eqn: main theorem diagonalization of FFFF}.

Importantly, we remark that in the optimisation given by Eq.~\eqref{eqn: SDP main thm} we are not optimising the SKR directly, therefore the theory of SDP does not guarantee that the values of $\pmb{\Lambda}$ found through this algorithm are the optimal for Eq.~\eqref{eqn: main theorem SKR}. Nevertheless, they are suitable values whose performance can be directly compared with the ones in Eq.~\eqref{eqn: AB bounds for lambda main thm}. 

In the next section we show a direct application of our analysis, together with the computation of $\pmb{\Lambda}$ through both Eqs.~\eqref{eqn: SDP main thm}-\eqref{eqn: AB bounds for lambda main thm}. 

{\section{Experimental characterisation and key rate simulation}\label{sec: experiment}

To illustrate the applicability of our method, we compute numerically the SKR based on the experimental characterisation of {two commercial SPDs}. In detail, we consider the {DEM} induced by the detectors response to different polarisation states for the incoming photons. We employ two ID Qube NIR Free Running single-photon avalanche diodes (SPADs) produced by {the company} ID Quantique \cite{httpsdet}. For both units, the nominal single-photon detection efficiency is set to $\eta_{th} = 25\%$ and the dead time to $\tau_{d}^{th} = 20 \ \mu$s.

While this choice of detectors is motivated by the availability of equipment in our lab, we remark that the polarisation-induced DEM of SPADs is widely recognised as small compared with the one of superconducting nanowire single-photon detectors (SNSPDs) \cite{feiFastAccurate2022}, for which it would be of interest to conduct an analogous analysis.

\subsection{Probing the polarisation-induced DEM} 

\begin{figure}
    \centering
    \includegraphics[width=0.4\linewidth]{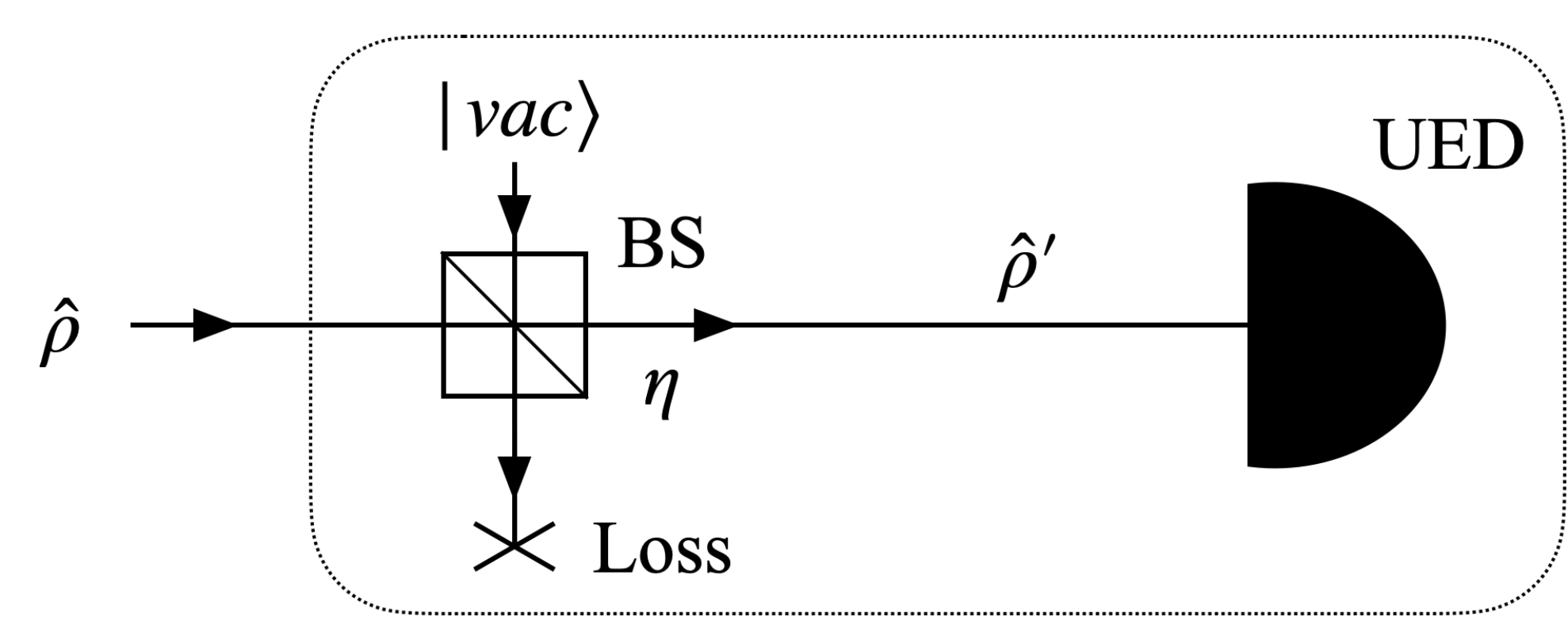}
    \caption{{Modelling of a realistic SPD (dotted box). {The input state $\hat{\rho}$ goes first through a beam splitter (BS) of transmittance $\eta$, which models the limited detector efficiency as the coupling of the input mode and the vacuum state ($\ket{vac}$) with a lost line.} Afterwards, the light is directed towards a unitary efficiency detector (UED) which has realistic dead time and dark count rate.}}
     \label{fig: Real det model}
\end{figure}

In this analysis we adopt the standard detector model depicted in Fig.~\ref{fig: Real det model}. In detail, the single-photon detection efficiency $\eta \le 1$  is modelled as the action of a beam splitter (BS) of equal trasmittance, placed before a unitary efficiency detector (UED). 
As in our case we are investigating polarisation-dependent detection efficiency, the action of the BS depends on the polarisation state of the incoming light.

Consider a flux of photons arriving at the UED at an average rate $R'_{in}$ and let $\tau_d$ denote the actual detector dead time. Following the analysis carried out in Ref.~\cite{knollRadiationDetection2000}, 
the average rate of total detected photons is given {by
\begin{equation}\label{eqn: function for fit detectors 1}
    R_{det} = \frac{R'_{in}}{1+R'_{in}\tau_d}.
\end{equation}}

{Suppose now that} the detector receives from the channel a train of single-photon states at a rate $R_{in, sp}$. {At each round, the input state $\hat{\rho}$ in Fig.~\ref{fig: Real det model} is given by $\hat{\rho}_{sp}=\dyad{1}$, with $\ket{1}$ being the single-photon Fock state. This means that} after the BS the state becomes
\begin{equation}\label{eqn: rho prime single photon exp}
    \hat{\rho}'_{sp} = \eta\dyad{1} + \pt{1-\eta}\dyad{vac},
\end{equation}
the latter term denoting the vacuum state. As a result, {it can be considered that} the UED receives photons at an incident rate $R'_{in} =\eta R_{in,sp} + R_{dark}$, since dark counts can be modelled as sporadic photons reaching the detector.
Substitution in Eq.~\eqref{eqn: function for fit detectors 1} yields a model to characterise the single-photon detection efficiency and the dead time of the detector by measuring the value of $R_{det}$ for different incoming photon {rates and then} interpolating the experimental data. 

Crucially, performing this characterisation would require a reliable single-photon source. In our case, we adopt a standard laser which is gain-switched at a repetition rate $\nu = 80$ MHz and is successively attenuated to the single photon level. Note that since $\nu \gg 1/\tau^{th}_d$, the detector sees the incoming light as a continuous wave. {Nevertheless, by pulsing the laser we mitigate beating effects due to reflections in the optical line, which might induce fluctuations of the average intensity that ultimately alter the results. In fact, a laser operating in continuous-wave mode can suffer from ``cavity effects" due to imperfect fiber connections that can cause multiple weak reflections which interfere with the original signal from the source, thus leading to fluctuations in the optical power arriving at the detector. This effect is greatly minimized when the laser is pulsed, since due to gain-switching the pulses have an approximately uniformly random phase relation, and thus the average power at the detectors is more stable.}

{Adopting a pulsed laser source, the signal arriving at the detector at each round consists of a weak coherent pulse of intensity $\mu$. Going through the BS in Fig.~\ref{fig: Real det model}, which models the non-ideal efficiency, causes an attenuation of the intensity $\mu \to \eta\mu$, which means that after the BS the state of the pulse takes the form
\begin{equation}
    \hat{\rho}'_{ch} = \dyad{\sqrt{\eta\mu}e^{i\phi}} = e^{-\eta\mu} \sum_{n,m=0}^{\infty} \frac{\pt{\sqrt{\eta\mu}}^{n+m}e^{i\pt{n-m}\phi}}{\sqrt{n!m!}} \ketbra{n}{m},
\end{equation}
where the last expression corresponds to the state expansion in the Fock basis. This implies that the probability of not being in a vacuum state is given by
\begin{equation}
    \sum_{k=1}^{\infty}\expval{\hat{\rho}'_{ch}}{k} = 1 - e^{-\eta\mu} \overset{\eta\mu\ll1}{\approx} \eta\mu, 
\end{equation}}
where the final approximation holds in our case since, for all the data acquisition rounds we perform, the value of $\mu$ satisfies $\mu \le 0.1$. 
Therefore, have that the average rate of incoming photons to the UED is given by
\begin{equation}\label{eqn: function for fit detectors 2}
    R'_{in} \approx \eta\mu\nu + R_{dark} = \eta R_{in, ch} + R_{dark}, 
\end{equation}
where we have defined the average rate of photons sent to the detector in the case of coherent light {$R_{in, ch} := \mu\nu$}. For ease of notation, in the following we neglect the subscript $ch$.

\begin{figure}
    \centering
    \includegraphics[width=0.6\linewidth]{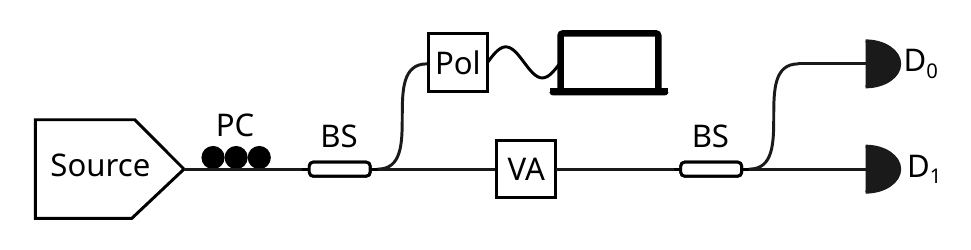}
    \caption{Experimental setup for the characterisation of the polarisation-dependent DEM of two detectors $D_0$ and $D_1$. Here PC, polarisation controller; Pol, polarimeter; BS, balanced beam splitter; VA, variable attenuator.}
     \label{fig: exp for characterisation}
\end{figure}

To measure the average count rate of the detectors while changing the rate and the polarisation state of the incoming photons we adopt the fiber-based scheme described in Fig.~\ref{fig: exp for characterisation}. 
Here, we study four polarisations states, namely horizontal ($\ket{H}$), vertical ($\ket{V}$), diagonal ($\ket{D} = \pt{\ket{H} + \ket{V}}/\sqrt{2}$) and circular-left ($\ket{L} = \pt{\ket{H} + i\ket{V}}/\sqrt{2}$), as they suffice in providing a full tomography of the operators $\{\opxprod{\hat{F}_s}\}_{s\in\{0,1\}}$. For each polarisation, we measure the average amount of detected photons per second over a $10$ s time window. The measured dark counts rates of the two SPDs over such time window are $930$ Hz and $630$ Hz, respectively. We note that since we do not employ polarisation-maintaining fiber, the polarisation state at the entrance of the SPDs might be, in all generality, different from the one measured by the polarimeter in Fig.~\ref{fig: exp for characterisation}, up to a unitary transformation. Hence, the basis adopted for the characterisation of the SPDs is the one of the polarisation states at the entrance of the first BS of Fig.~\ref{fig: exp for characterisation}.

For illustration purposes, we depict the results of this analysis for the case of horizontal and vertical polarisations in Fig.~\ref{fig: detector efficiency estimation plots}. The full results are reported in Table~\ref{tab: experiemntal results}. We remark that these values only serve the purpose of {explaining} the application of our analysis and, as such, performing a more extensive parameter estimation with high precision and small uncertainty goes beyond the scope of this paper.
Importantly, we observe a systematic decrease of both detectors efficiency for the polarisation state $\ket{H}$. This could be due to a joint DEM {of the two detectors, or to polarisation-dependent losses in the elements of the optical line that we employ in our characterisation setup (see Fig.~\ref{fig: exp for characterisation}). In detail, we have experimentally measured intensity fluctuations at the detectors entrance up to $5\%$ when changing the polarisation, which are compatible with the nominal polarisation-dependent loss in the BSs we adopt, which is up to $0.2$ dB according to the manufacturer.
Nevertheless, BSs are typically part of the measurement scheme and, in all generality, for the sake of proving the implementation security it is meaningful to consider the worst case scenario, that is, the one in which the discrepancy in efficiency we observe is due to polarisation-dependent DEM of the SPDs.}

\begin{figure*}
\hfill
\subfloat[Detector $D_0$.]{\includegraphics[width=0.49\columnwidth]{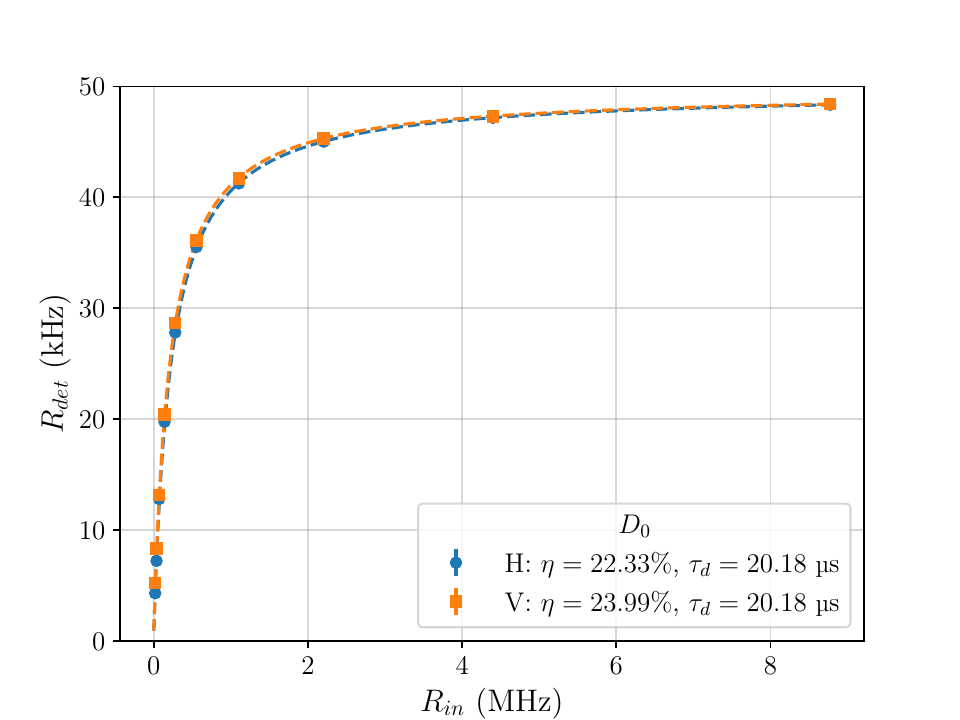}}\hfill
\subfloat[Detector $D_1$.]{\includegraphics[width=0.49\columnwidth]{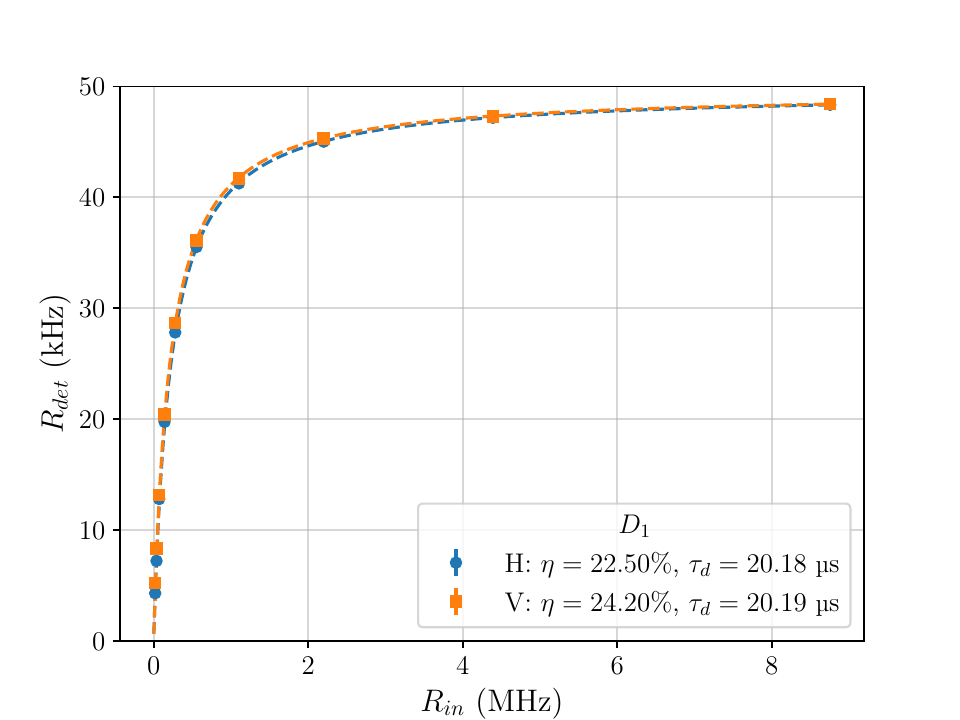}}    
\hfill
\caption{Experimental data for the characterisation of the {efficiency} of detectors $D_0$ (Fig.~\ref{fig: exp for characterisation}(a)) and $D_1$ (Fig.~\ref{fig: exp for characterisation}(b)), considering horizontal and vertical polarisation of the incoming light. We interpolate the experimental data (full markers) with the model given by Eqs.~\eqref{eqn: function for fit detectors 1}-\eqref{eqn: function for fit detectors 2} (dashed lines) {through the non-linear least squares method}. The efficiency $\eta$ and the dead time $\tau_d$ are then estimated as the optimal parameters of the interpolation. 
{We note that error bars associated to the experimental points in these plots, which correspond to the standard deviation of the measurements over the time window, are not quite visible, as they are approximately $1\%$ of the detected values.}}
\label{fig: detector efficiency estimation plots}
\end{figure*}

\begin{table}
    \centering
    \begin{tabular}{c|c|c||c|c||c|c||c|c|}
    \cline{2-9}
    & \multicolumn{2}{c||}{H} & \multicolumn{2}{|c||}{V} & \multicolumn{2}{|c||}{D} & \multicolumn{2}{|c|}{L} \\\cline{2-9}
    & $\eta$ & $\tau_d$ ($\mu$s) & $\eta$ & $\tau_d$ ($\mu$s) & $\eta$ & $\tau_d$ ($\mu$s) & $\eta$ & $\tau_d$ ($\mu$s) \\
    \hline 
         \multicolumn{1}{|c|}{$D_0$} & 22.33\% & 20.18 & 23.99\% & 20.18 & 23.78\% & 20.18 & 23.69\% & 20.18 \\
         \multicolumn{1}{|c|}{$D_1$} & 22.50\% & 20.18 & 24.20\% & 20.19 & 24.01\% & 20.19 & 23.86\% & 20.18 \\\hline
    \end{tabular}
    \caption{Results of the characterisation of the detection efficiency of two SPADs when changing the polarisation of the incoming light. The values in this table are obtained as the optimal parameters of the interpolation of the function in Eqs.~\eqref{eqn: function for fit detectors 1}-\eqref{eqn: function for fit detectors 2} over {the} experimental data.}
    \label{tab: experiemntal results}
\end{table}

It follows from Assumption (A5) that the efficiency of the detector $D_s$ for $s\in\pg{0,1}$ when the incoming photons are in a polarisation state $\ket{\sigma}\in\pg{\ket{H},\ket{V},\ket{D},\ket{L}}$ is given by
\begin{equation}
    \eta_{s}^{\sigma} := \expval{\opxprod{\hat{F}_s}}{\sigma}.
\end{equation}
Therefore, the analytical form of $\{\opxprod{\hat{F}_s}\}_{s\in\{0,1\}}$ can be computed from the experimental data as
\begin{equation}\label{eqn: how to compute oxprod_F from exp data}
    \opxprod{\hat{F}_s} = \mqty(\alpha_s&\beta_s\\\beta_s^*&\gamma_s), \quad \quad \text{ with } \quad \quad \left\{
    \begin{array}{c}
        \alpha_s = \eta_{s}^{H}  \\
        \gamma_s = \eta_{s}^{V}  \\
        \Re\pt{\beta_s} = \eta_{s}^{D} - \frac{1}{2}\pt{\eta_{s}^{H} + \eta_{s}^{V}} \\
        \Im\pt{\beta_s} = \frac{1}{2}\pt{\eta_{s}^{H} + \eta_{s}^{V}} - \eta_{s}^{L} .
    \end{array}
    \right.
\end{equation}
Combining Eq.~\eqref{eqn: how to compute oxprod_F from exp data} with the data {provided} in Table~\ref{tab: experiemntal results} gives the efficiency matrices of the two detectors:
{
\begin{equation}\label{eqn: F_0 example}
    \opxprod{\hat{F}_0} = \mqty(0.2233 &         0.0062-0.0052j \\
       0.0062+0.0052j& 0.2399       ),
\end{equation}
\begin{equation}\label{eqn: F_1 example}
    \opxprod{\hat{F}_1} = \mqty(0.2250  & 0.0066-0.0051j \\
       0.0066+0.0051j & 0.2420      ),
\end{equation}
}
while {the operators $\{\hat{F}_s\}_{s}$} for $s\in\{0,1\}$ can be found from the diagonalisation of the hermitian matrices in Eqs.~\eqref{eqn: F_0 example}-\eqref{eqn: F_1 example} as
\begin{equation}
    \opxprod{\hat{F}_s} = \hat{U}^{\dag}_s \hat{D}_s \hat{U}_s, \quad \quad \quad \quad \hat{F}_s = \hat{D}^{1/2}_s \hat{U}_s.
\end{equation}

\subsection{Secret key rate simulation}

To compute the SKR, we consider the case of a time-bin-encoding three-state protocol. To simulate the detection statistics, we focus on the case in which Eve manipulates the polarisation of the incoming light without affecting the signal state prepared by Alice. 
Formally, this corresponds to the case in which at every round
\begin{equation}
    \ket{\phi_{i}}_B\ket{\sigma_0}_T \xrightarrow{Eve} \ket{\phi_{i}}_B\ket{\sigma_E}_T,
\end{equation}
where $i\in\pg{0_Z,1_Z,0_X}$, $\ket{\sigma_0}_T$ denotes the (arbitrary) polarisation of the pulses sent by Alice and $\ket{\sigma_E}_T$ is the polarisation state set by Eve.
{For concreteness}, we choose the scenario in which Eve selects $\ket{\sigma_E}_T$ as the eigenstate corresponding to the minimum eigenvalue of $\opxprod{\hat{F}_1}$, effectively forcing {$D_1$} to operate at the minimum possible {detection} efficiency. In this case, the probability of each detector clicking when receiving a photon is given by
\begin{equation}\label{eqn: eta t s for sim}
    \eta_0^{\sigma_E} := \expval{\opxprod{\hat{F}_0}}{\sigma_E} = 22.01\%, \quad \quad \eta_1^{\sigma_E}:=\expval{\opxprod{\hat{F}_1}}{\sigma_E} = 22.16\%.
\end{equation}

{To account for SPFs, we} consider three possible values of the overlap of the flawed $Z$ states given by
\begin{equation}
    c_{01}^Z := \braket{\phi_{0_Z}}{\phi_{1_Z}}_B \in \pg{0.01,0.1,0.3},
\end{equation}
and study the case in which Alice sends her three states with equal {\textit{a priori}} probability $p_i=1/3$, while Bob measures {the incoming signals} in the $Z$ basis with probability $p_{Z_B} = 2/3$. The joint probability of Alice sending the state $i$ and Bob detecting $s_{\beta}$, with $s\in\pg{0,1}$ and $\beta\in\pg{X,Z}$, is then computed as
\begin{equation}
    {p_{s_{\beta},i}} = \frac{p_{{\beta}_B}}{3} \pq{\eta_{ch}  \abs{\braket{\phi_i}{s_{\beta}}}^2  \eta_s^{\sigma_E} \pt{1-\frac{p_{dark}}{2}} + p_{dark}\pt{1-\eta_{ch}  \abs{\braket{\phi_i}{s_{\beta}}}^2  \eta_s^{\sigma_E}}\pt{1-\frac{p_{dark}}{2}}} , 
\end{equation}
where $\eta_{ch} = 10^{-\alpha L/10}$ is the transmittance of a lossy channel of length $L$ with {attenuation coefficient} $\alpha=0.2$ dB/km for standard optical fiber. The value of $p_{dark}$ has been set to $10^{-6}$, corresponding to an upper bound on the dark count probability for these detectors in a QKD setup running at $1$ GHz.
Here we consider the case in which double clicks are randomly assigned to single clicks with equal probability 
\cite{gottesmanSecurityQuantum2004}.

\begin{figure*}
\hfill
\subfloat[Full view.]{\includegraphics[width=0.49\columnwidth]{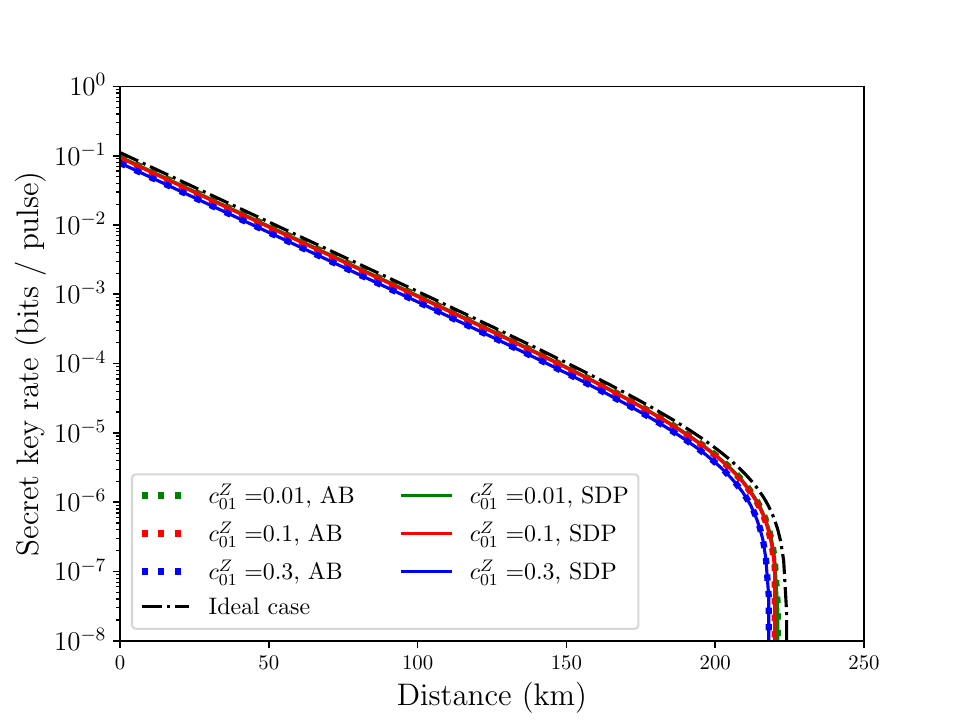}}\hfill
\subfloat[Zoomed view.]{\includegraphics[width=0.49\columnwidth]{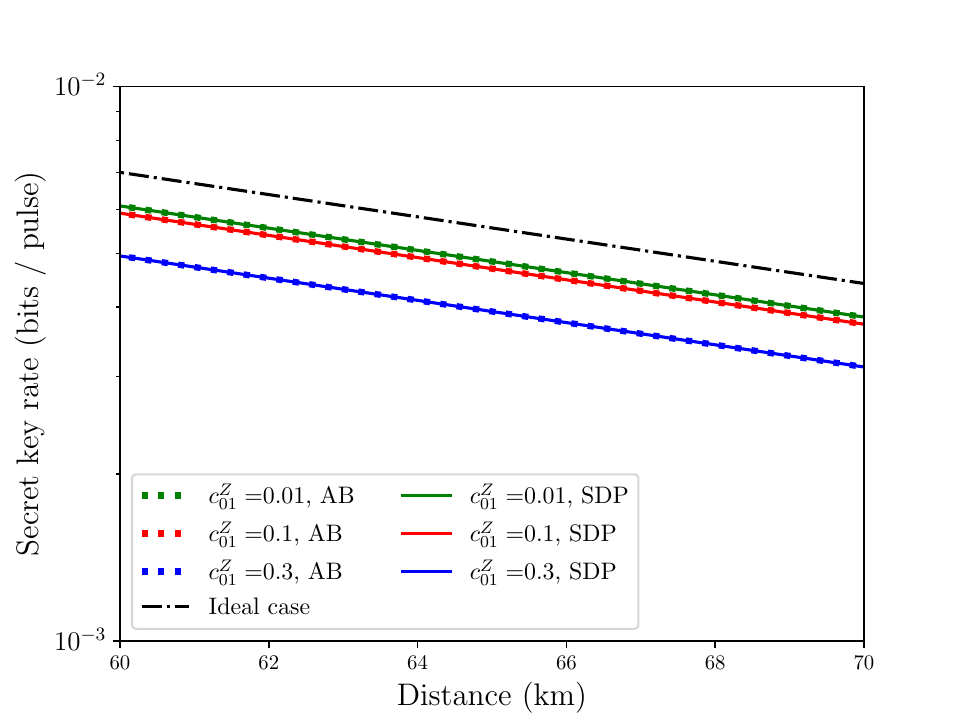}}    
\hfill
\caption{Lower bound on the asymptotic {SKR} for the three-state protocol with {SPFs and DEM corresponding to the detector matrices} reported in Eqs.~\eqref{eqn: F_0 example}-\eqref{eqn: F_1 example}. The results are obtained through the analysis in Sec.~\ref{sec: main theorem}, considering an {efficiency of the} error correction code $f=1.16$. We compare the performance of our approach adopting the values of $\pmb{\Lambda}$ provided by the analytical bounds (AB) in Eq.~\eqref{eqn: AB bounds for lambda main thm} (dashed lines) with the values {given by} the SDP in Eq.~\eqref{eqn: SDP main thm} (solid lines). The black {dash-dotted} line corresponds to the results for identical detectors working at the nominal efficiency (i.e., $\opxprod{\hat{F}_0} = \opxprod{\hat{F}_1} = 25\%$) and no SPFs. From the full picture (Fig.~\ref{fig: SKR plots example}(a)) it looks clear that the estimated {DEM}s does not affect {the performance} greatly. By looking in more detail (Fig.~\ref{fig: SKR plots example}(b)), we can {observe, as expected,} that higher SPFs correspond to lower SKR. Moreover, for the case we are considering, the results adopting the terms $\pmb{\Lambda}$ obtained thought analytical bounds or SDP coincide.}
\label{fig: SKR plots example}
\end{figure*}

With this choice of parameters we can compute the entries of the {terms $\pmb{\Tilde{q}}_s$} in Eq.~\eqref{eqn: main theorem LT q tilda}. To proceed to the calculation of the SKR {given by} Eq.~\eqref{eqn: main theorem SKR}, the only missing ingredient are the values of $\pmb{\Lambda}$. For these, we both solve the SDP in Eq.~\eqref{eqn: SDP main thm} and obtain the analytical bounds provided by Eq.~\eqref{eqn: AB bounds for lambda main thm}. The results are displayed in Fig.~\ref{fig: SKR plots example}. Critically, we observe that mismatches of the order of $5\%$ in the detectors efficiencies due to polarisation do not dramatically affect the performance of the protocol. Similarly, the role of {SPFs} is also mitigated, thanks to the loss-tolerant approach. For the specific case under investigation, we also {find} that the results of the analytical bounds and SDP coincide, although in all generality this might not always be the case.

\section{Discussion}\label{sec: discussion}
Results in Sec.~\ref{sec: main theorem} show directly how both SPFs and DEM {affect} the {SKR}. Importantly, we find that in case of identical detectors it holds $\lambda^{-}_s = \lambda^{+}_s = 1$ for $s\in\pg{0,1}$, and thus we retrieve the results of the original LT work \cite{tamakiLosstolerantQuantum2014}. In this {latter} case, experimental data from the actual protocol where Bob measures in the $X$ basis can be used to estimate the phase-error rate directly, due to the fact that {the basis-independent detection efficiency condition holds}.

We also note that in Eq.~\eqref{eqn: main theorem diagonalization of FFFF} the choice of which label (``$0$" or ``$1$") is assigned to the physical devices is completely arbitrary, hence one should choose the labeling that maximises Eq.~\eqref{eqn: main theorem SKR}.

Crucially, while the non-orthogonality of the $Z$ states directly affects the SKR, the deviation of $\ket{\phi_{0_X}}$ from $\ket{0_X}$ can be arbitrarily large, given that the state $\ket{\phi_{0_X}}$ is perfectly characterised. This is due to the fact that for the LT method the role of this state is only to help in the definition of the {terms $\pmb{\Tilde{q}}_s$} in Eq.~\eqref{eqn: main theorem LT q tilda}. In this respect, to achieve positive SKR Alice only needs to be able to prepare three pure states, select the most orthogonal ones as her key ($Z$) states and adopt the last one as trial state to compute the action of the channel on her signals.

Finally, a remark on the physical intuition behind the computation of the bounds in Eqs.~\eqref{eqn: main thm bound p_s}-\eqref{eqn: main thm bound e_p}. The potential of the LT approach lays in the fact that events in which Alice selects the $Z$ basis and Bob measures the incoming signals in the $X$ basis, which are typically discarded, can be actually used to characterise the channel action on the traveling systems and, therefore, help to estimate the phase-error rate. In {the} presence of DEM, this cannot be done in a straightforward way, as different detection probabilities in the two bases make impossible to directly relate the statistics of Bob's $Z$ and $X$ {basis} measurements. The role of the Procrustean virtual filter is precisely to remove this imbalance of the detectors, such that a scenario in which the detectors are equal can be restored in the virtual scheme. From here, operational inequalities such as the one in Eq.~\eqref{eqn: summary op bound for CdagC} can be employed to relate the statistics of Bob's virtual $X$ outcomes to his actual $X$ measurements, which he experimentally performs. This concept is illustrated in Fig.~\ref{fig: workflow} and further discussed in Appendix~\ref{sec: receiver}.

\section{Conclusions}\label{sec: conclusions}
In this paper, we have provided {a} security proof {for} the loss-tolerant {(LT)} QKD protocol, incorporating imperfections in both the source and measurement devices. Specifically, we {have considered a setup} subject to both state preparation flaws {(SPFs)} and detection efficiency mismatch {(DEM)}, for which the efficiency of the detectors varies according to several degrees of freedom of the received pulses, such as arrival time, frequency, polarisation, and spatial modes. 
{Under} assumptions on realistic SPFs and a experimentally-characterised polarisation-dependent DEM for commercial {single-photon} detectors, {our security proof shows that small} differences in the detectors efficiency do not affect {much} the key generation rate. In this respect, our works proves that LT QKD can be applied to setups with slightly unbalanced {detection efficiency} with minimal penalty.

We conclude this paper with some open questions. 
{We first remark for future reference that a more comprehensive analysis might benefit of relaxation of some assumptions we made in Sec.~\ref{sec: assumptions}, particularly the restriction of the information-carrying system to a qubit space for both Alice (Assumption (A1)) and Bob (Assumption (A4)). Moreover, as in this work we have only considered SPFs}, it is relevant to extend our security proof including also other major {source} imperfections, such as correlations among emitted pulses {and side-channels} \cite{pereiraQuantumFlawedLeaky2019,pereiraQuantumKey2020, zapateroSecurityQuantum2021, curras-lorenzoSecurityQuantum2024, curras-lorenzoSecurityFramework2023}.
{Finally,} our work provides a security proof in the asymptotic regime, where the number of emitted pulses is assumed to be infinite. {From here,} it will be of interest to conduct a security analysis for finite lengths. {In this latter case,} it should be noted that, as mentioned in the discussion leading to Eq.~\eqref{eqn: sum_l q as function of N}, it is necessary to consider probability trials across the entire train of emitted pulses. This supports the use of Kato's inequality \cite{katoConcentrationInequality2020}, which provides better convergence for finite lengths with respect to {Azuma's inequality \cite{azumaWeightedSums1967}.}

\section{Acknowledgements}
The authors thank G. Currás-Lorenzo, M. Pereira, \'A. Navarrete, X. Sixto, D. Rusca and V. Makarov for insightful discussions. 
The work of A. Marcomini is fully funded by the Marie Sklodowska-Curie Grant No. 101072637
(Project Quantum-Safe-Internet).
Support from the Galician Regional Government (consolidation of Research Units:
AtlantTIC), the Spanish Ministry of Economy and Competitiveness (MINECO), the
Fondo Europeo de Desarrollo Regional (FEDER) through the grant No.
PID2020-118178RB-C21, MICIN with funding from the European Union
NextGenerationEU (PRTR-C17.I1) and the Galician Regional Government with own
funding through the “Planes Complementarios de I+D+I con las Comunidades
Autónomas” in Quantum Communication, and the European Union’s Horizon Europe
Framework Programme under the project “Quantum Security Networks Partnership”
(QSNP, grant agreement No. 101114043) is greatly acknowledged. A. Mizutani is partially supported by JSPS KAKENHI Grant Number JP24K16977. K. Tamaki acknowledges support from JSPS KAKENHI Grant Number 23H01096.

\appendix
\section{Setup model}\label{sec: setup model}
In this Appendix we present a brief description of the model we use to describe the transmitter and receiver components of the QKD system considered, relating {them} to the assumptions reported in Sec.~\ref{sec: assumptions}. The notation we introduce here is later used in Appendix~\ref{sec: result proof} for the proof of our results.

\subsection{Transmitter}\label{sec: transmitter}
In a three-state {QKD scheme}, let $\ket{\phi_{0_Z}}$,  $\ket{\phi_{1_Z}}$ and $\ket{\phi_{0_X}}$ denote the flawed {quantum} states prepared by Alice (which can be in principle different from the perfect $\ket{0_Z}$, $\ket{1_Z}$ and  $\ket{0_X}$ states). These states are illustrated in Fig.~\ref{fig: flaweb Bloch sphere}. For later convenience, we make a choice of the axes of the Bloch sphere in such a way that all states share the same $Y$ coordinate and $\ket{\phi_{0_Z}}$ and $\ket{\phi_{1_Z}}$ are symmetric with respect to the $X$ axis. Note that this is always possible, {given} that the Bloch vectors of the flawed states are not all parallel (visually, this means that these states form a non-degenerate triangle on a plane cutting the Bloch sphere). Here we focus on the particular case of states laying on the $XZ$ plane of the sphere, since any set of states sharing the same $Y$ coordinate can be uniformly lifted to this plane through a filter $q\dyad{0_Y} + \pt{1-q}\dyad{1_Y}$ with $0\le q <1$. This ensures that $\braket{\phi_{0_Z}}{\phi_{1_Z}} \in \mathbb{R}$.

Therefore, the $Z$ states prepared by Alice for system $B$ can be written as
\begin{equation}\label{eqn: phi 0z as func of theta}
    \ket{\phi_{a_Z}}_B = \cos\pt{\frac{\theta - a \pi}{2}}\ket{0_Z}_B + \pt{-1}^a \sin\pt{\frac{\theta - a \pi}{2}}\ket{1_Z}_B,
\end{equation}
for $a=0,1$ and an angle $\theta \in (-\pi,\pi]$ (see Fig.~\ref{fig: flaweb Bloch sphere}), which typically satisfies $\abs{\theta}\ll1$ in practical implementations.

Equivalently, we can consider Alice preparing the joint states 
\begin{equation}\label{eqn: Psi_Z in Z basis}
    \ket{\Psi_Z}_{A'B} = \frac{1}{\sqrt{2}}\pt{\ket{0_Z}_{A'}\ket{\phi_{0_Z}}_{B} +  \ket{1_Z}_{A'}\ket{\phi_{1_Z}}_{B}},
\end{equation}
\begin{equation}\label{eqn: Psi_X in Z basis}
    \ket{\Psi_X}_{A'B} = \ket{0_X}_{A'}\ket{\phi_{0_X}}_{B},
\end{equation}
for a two-qubit system $A'B$. She then keeps the system $A'$ in her lab, while sending the system $B$ to Bob. When preparing $\ket{\Psi_Z}_{A'B}$, by measuring $A'$ in the $Z$ basis she eventually selects the state of system $B$. 

Now, let the virtual $X$ states of system $B$ be defined as the balanced superposition of the actual $Z$ states, namely 
\begin{align}
    \ket{\phi^{virt}_{0_X}}_B &:= \frac{\ket{{\phi}_{0_Z}}_B + \ket{{\phi}_{1_Z}}_B}{\norm{\ket{{\phi}_{0_Z}}_B + \ket{{\phi}_{1_Z}}_B}} \equiv \ket{0_X}_B, \nonumber \\
    \ket{\phi^{virt}_{1_X}}_B &:= \frac{\ket{{\phi}_{0_Z}}_B - \ket{{\phi}_{1_Z}}_B}{\norm{\ket{{\phi}_{0_Z}}_B - \ket{{\phi}_{1_Z}}_B}} \equiv \ket{1_X}_B.
    \label{eqn: definition of virtual states}
\end{align}
We notice that {these} virtual $X$ states coincide with the true $\hat{X}$ eigenstates, due to the fact that the flawed $Z$ states are symmetric with respect to the $X$ axis of the Bloch sphere.
Since $\ket{a_Z}_{A'} = \pt{\ket{0_X}_{A'} + \pt{-1}^a\ket{1_X}_{A'}}/\sqrt{2}$ for $a\in\pg{0,1}$, the state
$\ket{\Psi_Z}_{{A'}B}$ in Eq.~\eqref{eqn: Psi_Z in Z basis} can be rewritten as 
\begin{equation}\label{eqn: psi_Z in X basis}
    \ket{\Psi_Z}_{{A'}B} = \sqrt{p_{0_X^{virt}}}\ket{0_X}_{A'}\ket{\phi_{0_X}^{virt}}_{B} + \sqrt{p_{1_X^{virt}}}\ket{1_X}_{A'}\ket{\phi_{1_X}^{virt}}_{B},
\end{equation}
where $p_{0_X^{virt}}$ ($p_{1_X^{virt}}$) denotes the probability of Alice sending {Bob} the virtual state $\ket{\phi_{0_X}^{virt}}_{B}$ ($\ket{\phi_{1_X}^{virt}}_{B}$) {in the virtual scenario}. These probabilities are given by
\begin{equation}\label{eqn: prob of sending 0x virt}
\begin{gathered}
    p_{0_X^{virt}} = \frac{\norm{\ket{{\phi}_{0_Z}}_B + \ket{{\phi}_{1_Z}}_B}^2}{4} = \frac{1+\braket{\phi_{0_Z}}{\phi_{1_Z}}_B}{2}, \\
    p_{1_X^{virt}} = 1 - p_{0_X^{virt}}.
\end{gathered}
\end{equation}
Remarkably, Eq.~\eqref{eqn: psi_Z in X basis} highlights how measuring the virtual $X$ states Bob could predict the outcome of an $X$ measurement on Alice's system ${A'}$, when she prepares $\ket{\Psi_Z}_{{A'}B}$. This is the core aspect of Koashi's security proof {based on complementarity} \cite{koashiSimpleSecurity2009}. Moreover, although in practical applications Alice sends {Bob the states} $\ket{{\phi}_{0_Z}}_B$ and $\ket{{\phi}_{1_Z}}_B$ with equal {\textit{a priori}} probability in the key rounds, the probability of sending the virtual $X$ states are generally different (as clearly visible from Fig.~\ref{fig: flaweb Bloch sphere}, where the state $\ket{\phi_{0_X}^{virt}}$ is favored). 

\begin{figure}
    \centering
    \includegraphics[width=0.5\textwidth]{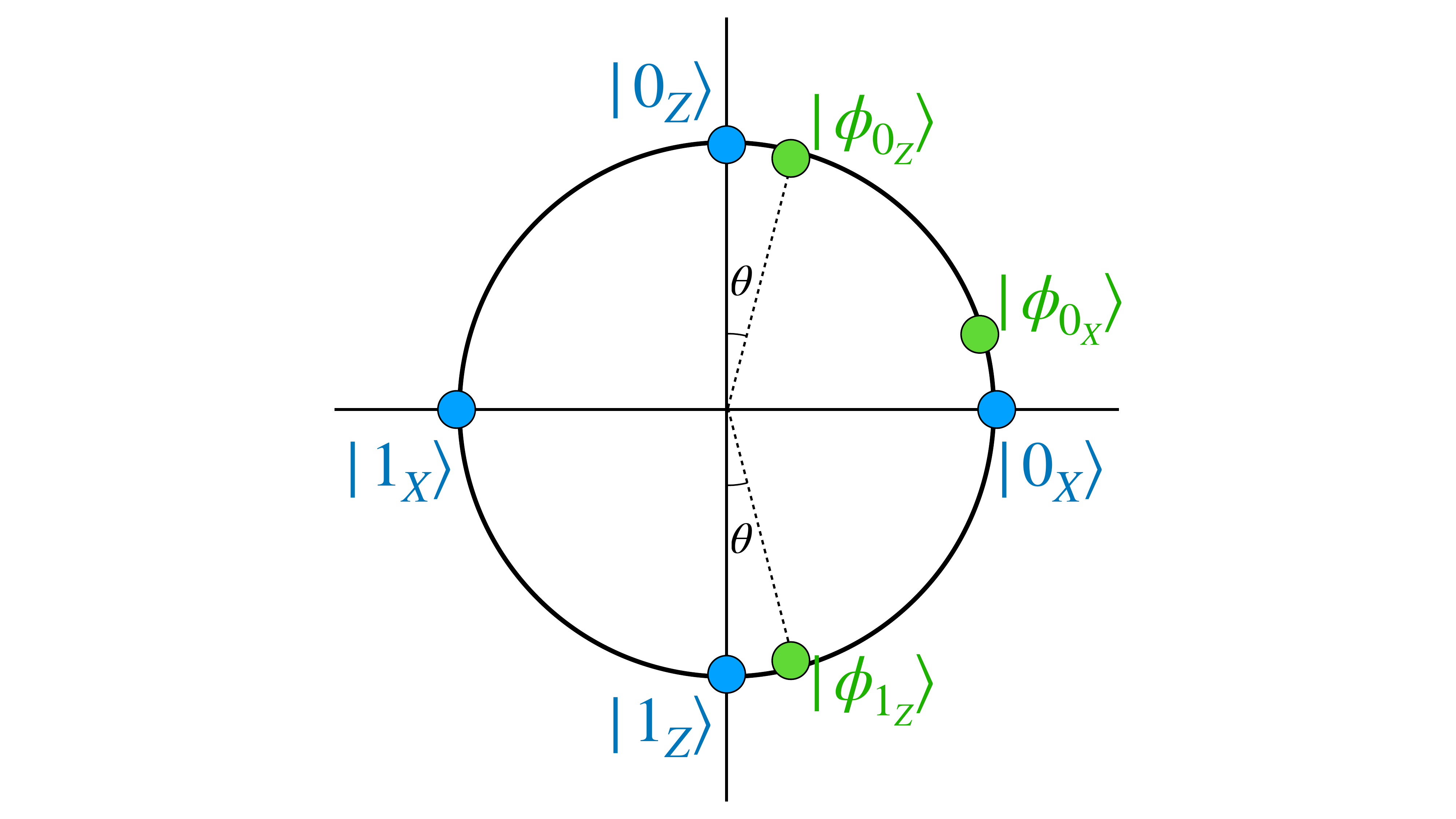}
    \caption{Illustration of the flawed {quantum} states $\ket{\phi_i}_{i\in\pgt{0_Z,1_Z,0_X}}$ for the $B$ system prepared by Alice (in green). Due to imperfections in the state preparation process, they differ from the eigenstates of the $\hat{X}$ and $\hat{Z}$ Pauli matrices (in blue). The choice of the $X$ and $Z$ axes of the Bloch sphere is done in such a way that the states $\ket{\phi_{0_Z}}$ and $\ket{\phi_{1_Z}}$ are symmetric with respect to the $X$ axis, i.e. they are rotated by the same angle $\theta$ from their ideal reference states.}
    \label{fig: flaweb Bloch sphere}
\end{figure}

As further discussed in Appendix~\ref{sec: result proof}, it is meaningful to study the evolution through the channel of all these states in the protocol, with a specific focus on the detection statistics in the $X$ basis for the actual states and in the $Z$ basis for the virtual states.   
{For this, we define the virtual, equivalent scheme introduced in Sec.~\ref{sec: protocol}. In this {virtual} scheme, an auxiliary shield system $A$ is adopted to describe a total of six different configurations, each corresponding to a specific state sent by Alice and a measurement choice for Bob. This description allows us to use the same notation for the analysis of the actual states sent and the one of the virtual $X$ states in Eq.~\eqref{eqn: definition of virtual states}, which are not actually sent.
In doing so, we have that the resulting state over the joint system $AB$ is in the form introduced in Eq.~\eqref{eqn:     varphi_AB},
and measuring in the orthonormal basis $\pg{\ket{c}_{A}}_{c\in\pg{1,\ldots,6}}$ of the shield system allows to select both a state $\ket{\phi^{\pt{c}}}_B$ and a measurement basis for Bob.
A summary of the virtual scheme is illustrated in Fig.~\ref{fig:fig_scheme_c_states}, while the states $\ket{\phi^{\pt{c}}}_B$, their corresponding probability and measurement basis are reported in Table~\ref{tab: virtual protocol recap}.} 

\begin{figure}[!ht]
    \centering
    \includegraphics[width=.6\textwidth]{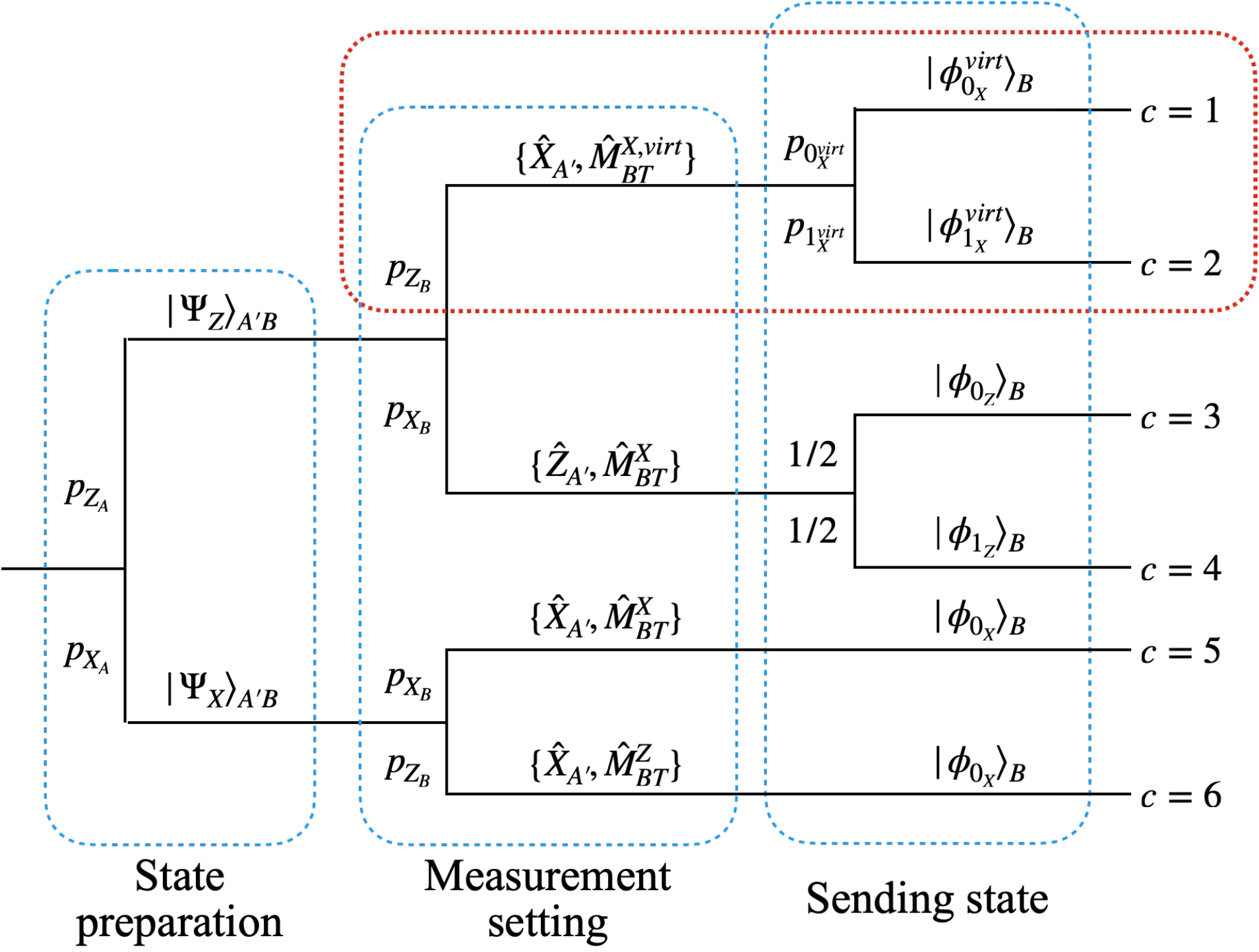}
    \caption{{This diagram illustrates the six scenarios corresponding to the states $\pgt{\ket{c}_A}_{c\in\pg{1,\ldots,6}}$ in the virtual protocol introduced in Sec.~\ref{sec: protocol} and explains the quantities in the summary provided in Table~\ref{tab: virtual protocol recap}. 
    Proceeding from left to right, Alice prepares the entangled states $\ket{\Psi_Z}_{A'B}$ and $\ket{\Psi_X}_{A'B}$ {given by} Eqs.~\eqref{eqn: Psi_Z in Z basis}-\eqref{eqn: Psi_X in Z basis} with probability $p_{Z_A}$ and $p_{X_A}$, respectively. Then, by performing a projective measurement $\hat{Z}_{A'}$ ($\hat{X}_{A'}$) in the $Z$ ($X$) basis, she post-selects the sending state of the $B$ system she is sending to Bob. For later use, here we already distinguish the possible measurement basis choices of Bob, including their probabilities in the definitions of the scenarios described by $c\in\pg{1,\ldots,6}$. From Assumption (A5) in Sec.~\ref{sec: assumptions}, Bob's measurements act on a joint system $BT$, being system $T$ the one describing the efficiency-affecting mode.
    The red box refers to the case of Bob performing a virtual $X$ basis measurement to predict the outcomes Alice would obtain by performing a measurement $\hat{X}_{A'}$ in the sifted key rounds ($c\in\pg{1,2}$). The phase-error rate is defined in terms of these outcomes. As the virtual measurement involves the sifted key rounds, the probability of performing the virtual measurement is the same of Bob performing an actual $Z$ measurement. 
    The cases for which $c\in\pg{3,4,5}$ correspond to the actual $X$ measurement rounds, while the case labelled by $c=6$ is not useful for our analysis and is considered for normalisation purposes only.}} 
    \label{fig:fig_scheme_c_states}
\end{figure}
\begin{table}[h!]
    \centering
    \begin{tabular}{|c|c|c|c|}\hline
    $c$ & $\ket{\phi^{\pt{c}}}_B$ & $p_c$ & Bob's basis \\ \hline
    $1$ & $\ket{\phi^{virt}_{0_X}}_B$ & $p_{0_X^{virt}}p_{Z_A}p_{Z_B}$ & virtual $X$ \\
    $2$ & $\ket{\phi^{virt}_{1_X}}_B$ & $p_{1_X^{virt}}p_{Z_A}p_{Z_B}$ & virtual $X$ \\ 
    $3$ & $\ket{{\phi}_{0_Z}}_B$ & $ p_{Z_A} p_{X_B} /2 $ & $X$ \\
    $4$ & $\ket{{\phi}_{1_Z}}_B$ & $ p_{Z_A} p_{X_B} /2 $ & $X$ \\
    $5$ & $ \ket{{\phi}_{0_X}}_B$ & $p_{X_A} p_{X_B}$ & $X$ \\ 
    $6$ & $ \ket{{\phi}_{0_X}}_B$ & $p_{X_A} p_{Z_B}$ & $Z$ \\ \hline
    \end{tabular}
    \caption{Emitted states $\ket{\phi^{\pt{c}}}_B$ and associated probabilities $p_c$ in the virtual protocol. 
    The quantity $p_{a_X^{virt}}$ {for $a\in\pg{0,1}$ is the probability of {Alice} sending the virtual state $\ket{{\phi}_{a_{X}}^{virt}}_B$ and is given by Eq.~\eqref{eqn: prob of sending 0x virt}, while $p_{\gamma_A}$ ($p_{\gamma_B}$) denotes the probability of Alice (Bob) selecting the basis $\gamma\in\pg{X,Z}$ to prepare (measure) the quantum states.}}
    \label{tab: virtual protocol recap}
\end{table}

Following the model introduced in Ref.~\cite{fungSecurityProof2009}, we consider an additional system $T$ which allows to characterise the nature of the efficiency mismatch. Intuitively, one can for example think of states $\pg{\ldots,\ket{-1}_T, \ket{0}_T, \ket{1}_T,\ldots}$ representing delays in the time of arrival of the single-photon to Bob's detector, or $\ket{\sigma}_T$ denoting a polarisation state for the incoming light as studied in Sec.~\ref{sec: experiment}, although this analysis applies to arbitrary mismatch-inducing modes. We shall introduce this system $T$ in a trivial configuration $\ket{0}_T$, allowing Eve to fully act on it in the channel. As a result, at each round of the protocol the joint state of the three systems $A$, $B$ and $T$ at the output of the transmitter is

\begin{equation}\label{eqn: varphi_ABT}
    \ket{\varphi}_{ABT} = \ket{\varphi}_{AB} \otimes \ket{0}_T,
\end{equation}
with $\ket{\varphi}_{AB}$ given by Eq.~\eqref{eqn:     varphi_AB}.

\subsection{Receiver}\label{sec: receiver}

{The operators $\opxprod{\hat{F}_s}$ acting on system $T$ that appear in Eq.~\eqref{eqn: full BT POVM definitions in assumptions}} represent the generalised efficiency of the two detectors \cite{fungSecurityProof2009}. {Precisely,} the quantity $\expval{\opxprod{\hat{F}_s}}{\gamma}_T$ represents the probability of the detector {$D_s$} clicking when the system $T$ is in a state $\ket{\gamma}$.
Note that Assumption (A3) in Sec.~\ref{sec: assumptions} implies that the operators $\opxprod{\hat{F}_s}$ are the same in the $Z$ and $X$ basis, as the same physical device used to detect $s_X$ is also employed to measure $s_Z$ (thus having the same {detection} efficiency).

The measurements {operators} $\hat{M}^{s_{\beta}}_{BT}$ introduced in Eq.~\eqref{eqn: full BT POVM definitions in assumptions} can be equivalently modelled as the result of a filter
\begin{equation}\label{eqn: Z filter}
    \hat{Q}_{\beta} = \dyad{0_{\beta}}_B \otimes \pt{\hat{F}_0}_T + 
                \dyad{1_{\beta}} \otimes \pt{\hat{F}_1}_T,
\end{equation}
followed by a projection ${\ketbra{s_{\beta}}_B \otimes \id_T}$ on the system $B$. In fact, we have that 
\begin{equation}
    \hat{M}^{s_{\beta}}_{BT} = \hat{Q}_{\beta}^{\dag} \pt{\ketbra{s_{\beta}}_B \otimes \id_T}\hat{Q}_{\beta}.
\end{equation}
This is relevant because the filtering discriminates the ``successful" and ``failure" bit measurement outcome, while the following projective measurement simply assigns the bit value {to successful events}. This is due to the fact that the operator associated to failed filtering for $\hat{Q}_{\beta}$ is indeed $\sqrt{\hat{M}^{{fail}_{\beta}}_{BT}}$.
In this respect, the sifted key rounds are {those} where the filtering $\hat{Q}_Z$ is successful (note that Alice's measurements of her shield system are always successful). We summarise the two equivalent descriptions of Bob's POVMs in Fig~\ref{fig: POVMs Z and X}.

We conclude remarking that although our analysis applies to a system $T$ of arbitrary dimension, for all practical purposes it might be beneficial to consider it to be finite \cite{fungSecurityProof2009}. In fact, Bob could, in principle, apply some mode-correction filtering at the entrance of his lab, effectively limiting the dimensionality of possible {input} modes. For example, when the system $T$ models shifts in the time of arrival of the signals, one can consider a finite number of discrete shifts by introducing a Gaussian filtering operation at the entry of the lab \cite{fungSecurityProof2009}. Moreover, if only a partial characterisation of the efficiency operators is available, one {should} consider the worst case scenario minimising the SKR given by Eq.~\eqref{eqn: main theorem SKR} over the possible missing entries of $\opxprod{\hat{F}_s}$, which might cause the problem to be intractable for a system $T$ of infinite dimension.
Finally, note that while we restrict our analysis to a single efficiency-affecting mode, this approach can be generalised in a straightforward way to a scenario with several efficiency-affecting modes, simply by considering multiple systems $T_1, \ldots, T_n$ and defining efficiency matrices $\opxprod{\hat{F}_s}$ for each of them.

 \begin{figure}
     \centering
     \includegraphics[width=0.75\linewidth]{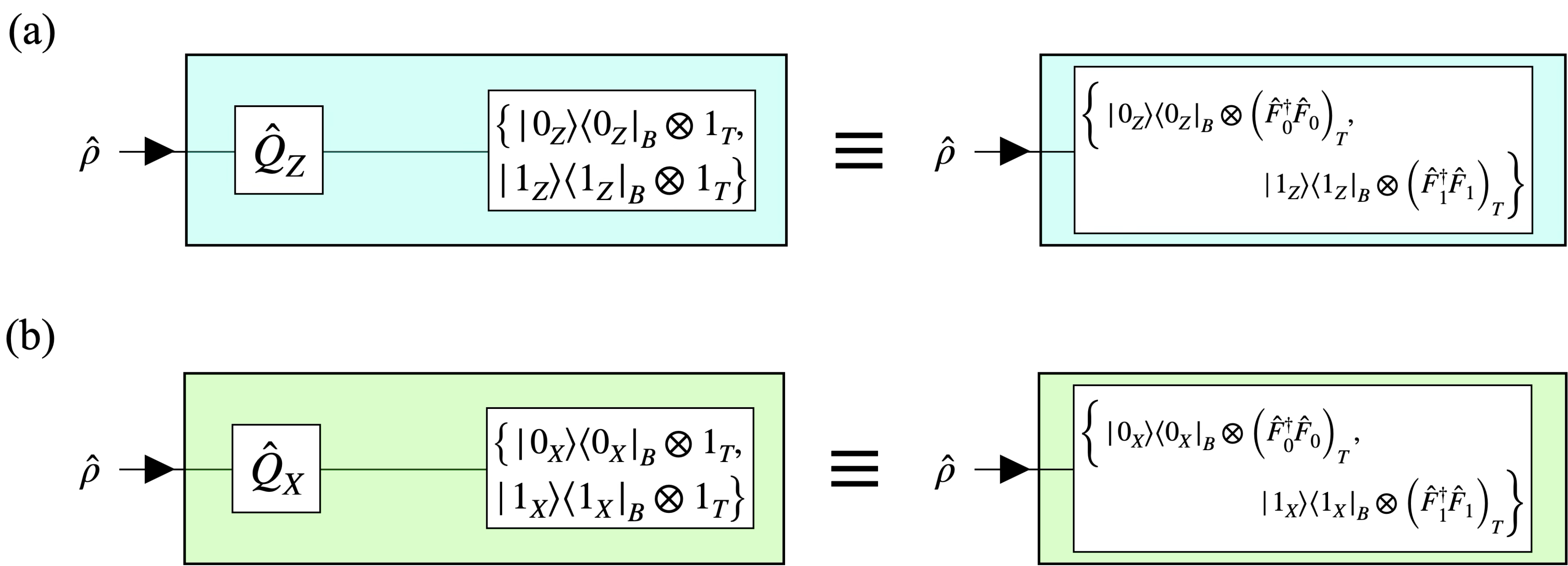}
     \caption{
     Bob's scheme for a successful measurement in (a) 
     the $Z$ basis and (b) the $X$ basis. For each figure, Bob's POVM elements corresponding to a successful measurement outcome (right-hand side) can be equivalently described by a filter $\hat{Q}_{\beta}$, {with} $\beta\in\pg{X,Z}$, which determines whether {or not the states $\hat{\rho}$ coming from the channel trigger a detection,} followed by projective measurements to assign the bit value (left-hand side). A failed measurement is equivalently described by the $\hat{M}^{s_{\beta}}_{BT}$ POVM element or the $\hat{Q}_{\beta}$ filter not succeeding.}
     \label{fig: POVMs Z and X}
 \end{figure}

\section{Security against coherent attacks}\label{sec: result proof}

In this Appendix we provide the full security proof against coherent attacks. We do so by first developing an analytical description of the way Bob's outcomes for the actual signal states {are} affected by Eve's attack and by his measurement schemes (Sec.~\ref{sec: Security analysis of the actually sent states}). From here, we extend the description to the case of virtual states, relating their statistics to {those} of the actual states (Sec.~\ref{sec: Security analysis of the virtual states}). We conclude the analysis by computing a lower bound on the SKR (Sec.~\ref{sec: Computing the secret key rate}) through an upper bound on the phase-error rate (Sec.~\ref{sec: bound ep epprime}).
The conceptual map of our approach is depicted in Fig.~\ref{fig: workflow}.

\begin{figure}
    \centering
    \includegraphics[width=0.9\linewidth]{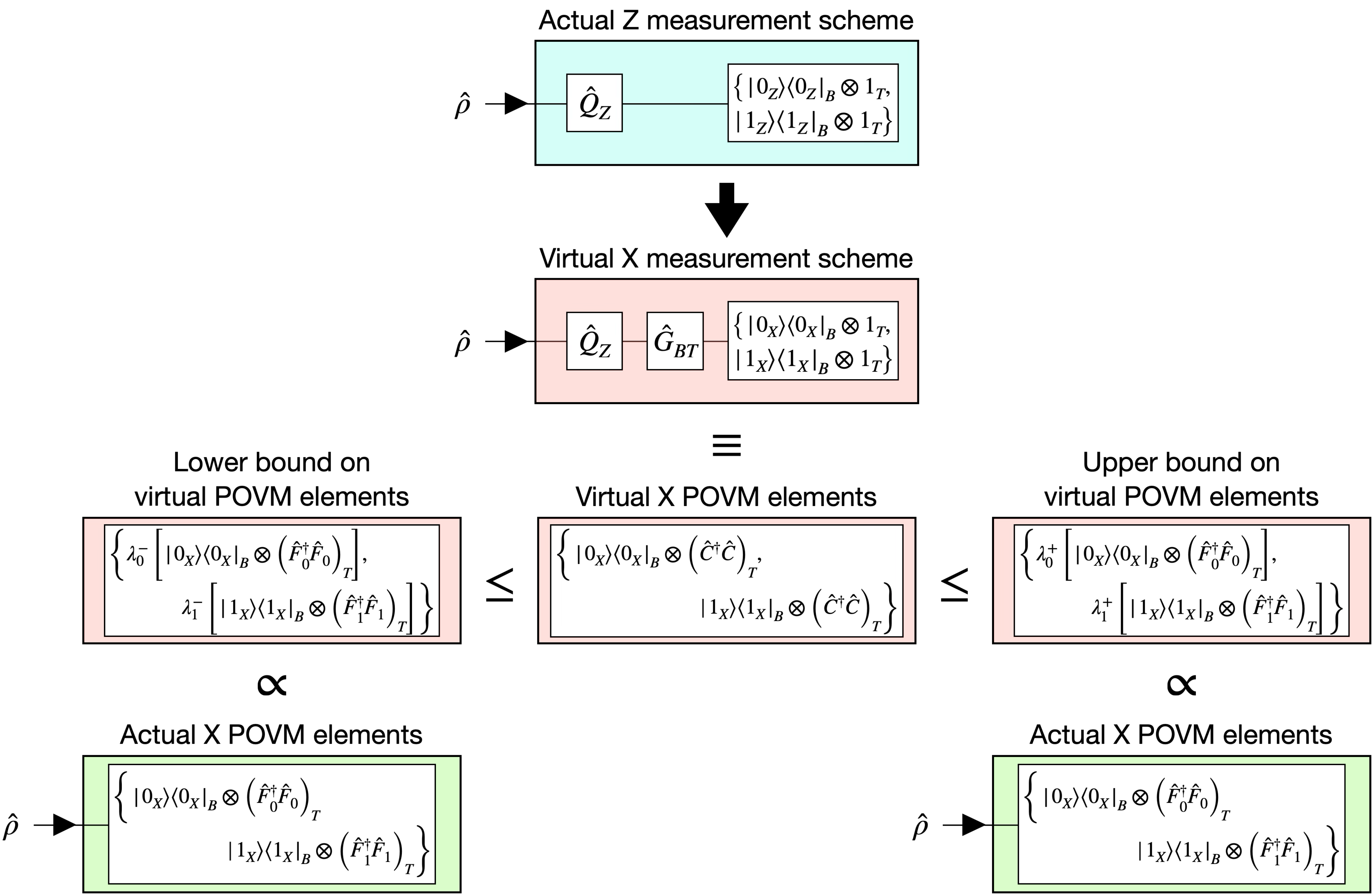}
    \caption{Workflow of the security proof provided in this paper. The boxes in this figure refer to either the actual measurements illustrated in Fig.~\ref{fig: POVMs Z and X}, or the virtual measurement reported in Fig.~\ref{fig: virtual POVM}. 
    For the sifted key rounds, we consider a virtual framework where Bob's actual $Z$ measurement is substituted by a virtual one. The latter is designed by first employing the same filter $\hat{Q}_Z$, such that the length of the sifted key in the actual and virtual schemes is preserved. Then, an additional virtual filter $\hat{G}_{BT}$ is applied, before a projective measurement in the $X$ basis is performed.
    By properly choosing the virtual filter $\hat{G}_{BT}$, we can balance the {DEM}, such that the efficiency associated to Bob's measurements on system $B$ is the same for every state of the system $T$. From here, we lower and upper bound these POVM operators so that each efficiency operator $\opxprod{\hat{F}_s}$ is coupled to the projector $\dyad{s_X}_B$ in the $X$ basis, for $s\in\pg{0,1}$. Crucially, these bound operators are now proportional to the measurements that Bob actually performs in the $X$ basis. As a result, we can relate Bob's virtual measurement results to the experimental data.}
    \label{fig: workflow}
\end{figure}

\subsection{Security analysis of the actual states}\label{sec: Security analysis of the actually sent states}
Let 
\begin{equation}
    \ket{\Phi}_{ABT} = \ket{\varphi^{\pt{<l}}}_{ABT}\ket{\varphi^{\pt{l}}}_{ABT}\ket{\varphi^{\pt{>l}}}_{ABT},
\end{equation}
denote the global state prepared by Alice during $N$ rounds of the protocol, where $\ket{\varphi^{\pt{<l}}}_{ABT}$ ($\ket{\varphi^{\pt{>l}}}_{ABT}$) denotes the full set of states of {the rounds that precede (follow) round $l$}. The $l-$th round state takes the form given by Eq.~\eqref{eqn: varphi_ABT}.
 
For coherent attacks, the action of Eve is represented by a joint unitary operator {$\hat{V}_{BTE}$ acting on all} the transmitted systems $BT$, together with Eve's ancillary system $E$, resulting in a state $\ket{\Psi}$ given by
\begin{equation}
    \ket{\Psi} = \hat{V}_{BTE} \ket{\Phi}_{ABT}\ket{0}_E
    = \sum_{i}  \hat{E}^{\pt{i}}_{BT}\ket{\Phi}_{ABT}\ket{i}_E.
\end{equation}

Let now
\begin{equation}
    \hat{O}_{ABT}^{\pt{l-1}} := \bigotimes_{u=1}^{l-1} \hat{M}^{c^u, s^u_{\beta}}_{ABT},
\end{equation}
where $\hat{M}^{c^u, s^u_{\beta}}_{ABT} := \dyad{c
}_{A_u} \otimes \hat{M}^{s_{\beta}}_{B_uT_u}$ is the Kraus operator associated to the measurement of the systems $A$, $B$ and $T$ at round $u$, yielding outcome $c$ for the shield system and $s\in\pg{0,1}$ for Bob's measurement in the basis $\beta\in\pg{X,Z}$. The joint state of systems $A$, $B$ and $T$ at round $l$ conditioned on the previous observations is then given by
\begin{equation}\label{eqn: rho cond on past introduction}
\begin{gathered}
    \hat{\rho}_{ABT}^{l|O^{\pt{l-1}}} = \sum_i \Tr_{\bar{l}}\pq{\hat{P}\pt{\hat{O}_{ABT}^{\pt{l-1}}\hat{E}^{\pt{i}}_{BT}\ket{\Phi}_{ABT}}}
    / p^{\pt{l}},\\
    p^{\pt{l}} = \Tr\pq{\sum_i \Tr_{\bar{l}}\pq{\hat{P}\pt{\hat{O}_{ABT}^{\pt{l-1}}\hat{E}^{\pt{i}}_{BT}\ket{\Phi}_{ABT}}}},
\end{gathered}
\end{equation}
where $\Tr_{\bar{l}}$ represents the trace over all systems but $A,B$ and $T$ at round $l$ and $\hat{P}\pt{\ket{ \psi }} := \dyad{ \psi }$. By letting $\pg{\ket{\vec{x}_{<l}}}$ ($\pg{\ket{\vec{x}_{>l}}}$) denote {an orthonormal} basis for the state of {the} rounds before (after) $l$, we can explicitly write the trace operator in Eq.~\eqref{eqn: rho cond on past introduction} to find
\begin{align}\nonumber 
    \hat{\rho}_{ABT}^{l|O^{\pt{l-1}}} \cdot p^{\pt{l}} &= 
    \sum_i \Tr_{\bar{l}}\pq{\hat{P}\pt{\hat{O}_{ABT}^{\pt{l-1}}\hat{E}^{\pt{i}}_{BT}\ket{\Phi}_{ABT}}}
    \\\nonumber
    &= 
    \sum_i \sum_{{\vec{x}_{<l}}}\sum_{\vec{x}_{>l}}\bra{\vec{x}_{<l}}\mel{\vec{x}_{>l}}{\hat{P}\pt{\hat{O}_{ABT}^{\pt{l-1}}\hat{E}^{\pt{i}}_{BT}\ket{\Phi}_{ABT}}}{\vec{x}_{>l}}\ket{\vec{x}_{<l}}_{ABT}
    \\\nonumber
    &= 
    \sum_i \sum_{{\vec{x}_{<l}},{\vec{x}_{>l}}}\hat{P}\pt{\hat{A}^{\pt{\vec{x}_{<l},\vec{x}_{>l}}}_{i;BT|O^{\pt{l-1}}}\ket{\varphi^{\pt{l}}}_{ABT}}
    \\
    &= 
    \sum_i \sum_{{\vec{x}_{<l}},{\vec{x}_{>l}}}\hat{P}\pt{\sum_c \sqrt{p_c}\ket{c}_{A}\hat{A}^{\pt{\vec{x}_{<l},\vec{x}_{>l}}}_{i;BT|O^{\pt{l-1}}}\ket{\phi^{\pt{c}}}_{B}\ket{0}_T},
    \label{eqn: first definitoon rho cond}
\end{align}
where we have defined
\begin{equation}\label{eqn: def o A terms}
    \hat{A}^{\pt{\vec{x}_{<l},\vec{x}_{>l}}}_{i;BT|O^{\pt{l-1}}} := \bra{\vec{x}_{<l}}\mel{\vec{x}_{>l}}{\hat{O}_{ABT}^{\pt{l-1}}\hat{E}^{\pt{i}}_{BT}}{{\varphi^{\pt{<l}}}}_{ABT}\ket{\varphi^{\pt{>l}}}_{ABT}.
\end{equation}
This operator acts jointly on the systems $B$ and $T$ at round $l$ and can be decomposed as
\begin{equation}
    \hat{A}^{\pt{\vec{x}_{<l},\vec{x}_{>l}}}_{i;BT|O^{\pt{l-1}}} = \sum_k \hat{A}^{\pt{\vec{x}_{<l},\vec{x}_{>l}}}_{i,k;B|O^{\pt{l-1}}} \otimes \hat{A}^{\pt{\vec{x}_{<l},\vec{x}_{>l}}}_{i,k;T|O^{\pt{l-1}}}
\end{equation}
over the two systems. For further reference, we define the (unnormalised) action of the {element $\hat{A}^{\pt{\vec{x}_{<l},\vec{x}_{>l}}}_{i,k;T|O^{\pt{l-1}}}$} on the system $T$ as
\begin{equation}\label{eqn: unnormalised gamma T}
    \ket{\gikl}_T := \hat{A}^{\pt{\vec{x}_{<l},\vec{x}_{>l}}}_{i,k;T|O^{\pt{l-1}}}\ket{0}_T,
\end{equation}
such that from Eq.~\eqref{eqn: first definitoon rho cond} it follows
\begin{equation}
    \hat{\rho}_{ABT}^{l|O^{\pt{l-1}}}  = \frac{1}{p^{\pt{l}}}
    \sum_{i} \sum_{{\vec{x}_{<l}},{\vec{x}_{>l}}}\hat{P}\pt{\sum_{c,k} \sqrt{p_c}\ket{c}_{A}\hat{A}^{\pt{\vec{x}_{<l},\vec{x}_{>l}}}_{i,k;B|O^{\pt{l-1}}}\ket{\phi^{\pt{c}}}_{B}\ket{\gikl}_T}.
\end{equation}

We are interested in computing the joint probabilities $P_{s_X,c|O^{\pt{l-1}}}$ of Bob obtaining outcome $s_X$ at round $l$ by measuring the system $B$ in the $X$ basis and Alice sending the $c-$th state $\ket{\phi^{\pt{c}}}_B$, conditioned on {all} the previous results. For the actual states measured in the $X$ basis (that is, when $c\in\pg{3,4,5}$), we have that
\begin{align}
    P_{s_X,c|O^{\pt{l-1}}} &= \Tr\pq{\pt{\dyad{c}_{A}\otimes \hat{M}^{s_{X}}_{BT}}\hat{\rho}_{ABT}^{l|O^{\pt{l-1}}}} \nonumber \\
    &= \frac{p_c}{p^{\pt{l}}}
    \sum_{i} \sum_{{\vec{x}_{<l}},{\vec{x}_{>l}}}\Tr\pq{{ \hat{M}^{s_{X}}_{BT}}\hat{P}\pt{\sum_k\hat{A}^{\pt{\vec{x}_{<l},\vec{x}_{>l}}}_{i,k;B|O^{\pt{l-1}}}\ket{\phi^{\pt{c}}}_{B}\ket{\gikl}_T}}
    \nonumber \\
    &= \frac{p_c}{p^{\pt{l}}} \sum_{i,k,k'} \sum_{{\vec{x}_{<l}},{\vec{x}_{>l}}}
    \mel{\giklp}{\opxprod{\hat{F}_s}}{\gikl}_T \nonumber \\
    & \quad \quad \quad \quad \quad \quad \quad \Tr\pq{\hat{A}^{\pt{\vec{x}_{<l},\vec{x}_{>l}}\dag}_{i,k';B|O^{\pt{l-1}}}\dyad{s_X}_B\hat{A}^{\pt{\vec{x}_{<l},\vec{x}_{>l}}}_{i,k;B|O^{\pt{l-1}}}\hat{P}\pt{\ket{\phi^{\pt{c}}}_B}} \nonumber \\
    &= p_c\Tr\pq{\hat{D}_{s_X|O^{\pt{l-1}}}\hat{P}\pt{\ket{\phi^{\pt{c}}}_B}},
    \label{eqn: P for coherent case LT+DEM}
\end{align}
where we have defined
\begin{equation}
    \hat{D}_{s_X|O^{\pt{l-1}}} := \frac{1}{p^{\pt{l}}}\sum_{i,k,k'}\sum_{{\vec{x}_{<l}},{\vec{x}_{>l}}} \mel{\giklp}{\opxprod{\hat{F}_s}}{\gikl}_T {\hat{A}^{\pt{\vec{x}_{<l},\vec{x}_{>l}}\dag}_{i,k';B|O^{\pt{l-1}}}}\dyad{s_X}_B\hat{A}^{\pt{\vec{x}_{<l},\vec{x}_{>l}}}_{i,k;B|O^{\pt{l-1}}}.
\end{equation} 

We can now proceed to decompose $\hat{P}\pt{\ket{\phi^{\pt{c}}}_B}$ in the Pauli basis. Let 
\begin{equation}\label{eqn: def sigma op as pauli matrices}
\pt{\hat{\sigma}_{\nohatid},\hat{\sigma}_{X},\hat{\sigma}_{Y},\hat{\sigma}_{Z}} = \pt{\id,\hat{X},\hat{Y},\hat{Z}}
\end{equation}
be the set of identity and Pauli matrices for system $B$. Then:
\begin{equation}\label{eqn: decomposition of Bloch state in terms of Pauli}
    \hat{P}\pt{\ket{\phi^{\pt{c}}}_B} = \frac{1}{2}\sum_{\omega \in \Omega} {W}^{\pt{c}}_{\omega} \hat{\sigma}_{\omega},  
\end{equation}
where $\Omega=\pg{\nohatid,X,Y,Z}$ and $\pmb{W}^{\pt{c}} := \pt{1,{V}^{\pt{c}}_X, {V}^{\pt{c}}_Y, {V}^{\pt{c}}_Z}^T$, being ${V}^{\pt{c}}_{t}$ for $t\in\pg{X,Y,Z}$ the entries of the Bloch vector of the state $\ket{\phi^{\pt{c}}}_B$. Here, ${W}^{\pt{c}}_{\nohatid}=1$. {Also, note} that we have ${V}^{\pt{c}}_Y = 0$ $\forall c$, as from Assumption (A1) all flawed states lay on the $XZ$-plane of the Bloch sphere. {Substituting Eq.~\eqref{eqn: decomposition of Bloch state in terms of Pauli}} in Eq.~\eqref{eqn: P for coherent case LT+DEM}, one finds
\begin{equation}\label{eqn: P_sx,c for coherent}
    P_{s_X,c|O^{\pt{l-1}}} = p_c\sum_{\omega \in \Omega} {W}^{\pt{c}}_{\omega} q_{s_X,\omega|O^{\pt{l-1}}},
\end{equation}
with
\begin{equation}\label{eqn: q definition}
    q_{s_X,\omega|O^{\pt{l-1}}} := \frac{1}{2} \Tr\pq{\hat{D}_{s_X|O^{\pt{l-1}}}\hat{\sigma}_{\omega}}.
\end{equation}

By means of Kato's inequalities \cite{katoConcentrationInequality2020}, {we have that} the sum of the probabilities in Eq~\eqref{eqn: P_sx,c for coherent} over all $N$ rounds is asymptotically close to the number of occurrences of the {corresponding events} in the protocol, that is
\begin{equation}\label{eqn: sum_l q as function of N}
    \sum_{l=1}^NP_{s_X,c|O^{\pt{l-1}}} = p_c\sum_{\omega \in \Omega} {W}^{\pt{c}}_{\omega} \sum_{l=1}^N q_{s_X,\omega|O^{\pt{l-1}}} \xrightarrow{N\to \infty} N_{s_X,c},
\end{equation}
where $N_{s_X,c}$ is the number of times Alice {sends} the state $\ket{\phi^{\pt{c}}}_B$ and Bob {obtains} the result $s_X$.
From the idea of the LT protocol, as long as the vectors $\pgt{\vec{W}^{\pt{c}}}_{c\in\pg{3,4,5}}$ of the flawed {quantum} states are linearly independent, inverting the set of {equations provided by Eq.~\eqref{eqn: sum_l q as function of N}} for $c\in\pg{3,4,5}$ allows to estimate the values of the terms $\sum_{l=1}^N q_{s_X,\omega|O^{\pt{l-1}}}$ for $\omega\in\pg{\nohatid,X,Z}$ from the experimental data \cite{tamakiLosstolerantQuantum2014}. This corresponds to the result in Eq.~\eqref{eqn: main theorem LT q tilda}, where $\forall \omega$ we {define 
\begin{equation}\label{eqn: qtilda def}
    \Tilde{q}_{s_X,\omega} := \lim_{N\to\infty}\frac{1}{N}\sum_{l=1}^N q_{s_X,\omega|O^{\pt{l-1}}},
\end{equation}
and consider in the asymptotic limit 
\begin{equation}\label{eqn: asymptotic definition of p_sx_c}
    p_{s_X,c} := \lim_{N\to\infty}\frac{N_{s_X,c}}{N}.
\end{equation}}

\subsection{Security analysis of the virtual states}\label{sec: Security analysis of the virtual states}

In the complementarity framework, the phase-error rate can be computed once we find the errors that Bob encounters in predicting Alice's virtual $X$ outcomes for the states they use to produce the sifted key.
Crucially, due to {the DEM}, the probability of successful detection in the $Z$ or $X$ basis for Bob is not the same. Therefore, to balance statistics, at round $l$ we must consider the state going through the $Z$ basis filter in Eq.~\eqref{eqn: Z filter}, namely
\begin{equation}\label{eqn: input state conditioned on Q_Z for coherent}
    \hat{Q}_Z\hat{\rho}_{ABT}^{l|O^{\pt{l-1}}}\hat{Q}_Z^{\dag} .
\end{equation}
See Fig.~\ref{fig: workflow} for reference.

To predict Alice's virtual $X$ outcomes, Bob is allowed to apply any virtual filter $\hat{G}_{BT}$ on the above states. The goal is to design a virtual filter that allows him to rely on the information available from {the} experimental data. Ideally, the best option would be a virtual filter satisfying $\hat{G}_{BT}\hat{Q}_Z = \hat{Q}_X$. This would imply that for all events for which a conclusive outcome can be asserted in the $Z$ basis, we can obtain a virtual $X$ measurement conclusive event as well, which we can relate to the actual statistics in the $X$ basis that Bob observes in his experiment. Crucially, due to the DEM this is not possible. 

In this paper we adopt the filter proposed in Ref.~\cite{fungSecurityProof2009}, which is given by
\begin{equation}\label{eqn: virtual filter introduction}
    \hat{G}_{BT} = \dyad{0_Z}_B\otimes \hat{C}\hat{F}_0^{-1} + \dyad{1_Z}_B\otimes \hat{C}\hat{F}_1^{-1},
\end{equation}
followed by a projective measurement {on $\dyad{s_X}_B \otimes \id_T$}.
This corresponds to the definition of the following virtual POVMs acting on the state incoming from the channel
\begin{equation}\label{eqn; definition of M virt}
    \hat{M}^{s_X,virt}_{BT} = \hat{Q}_Z^{\dag}\hat{G}_{BT}^{\dag}\pt{\dyad{s_X}_B \otimes \id_T}\hat{G}_{BT}\hat{Q}_Z.
\end{equation}
Importantly, the $\hat{C}$ matrix is chosen so to guarantee that $\hat{G}_{BT}$ is a valid filtering operation. 
Moreover, note that
\begin{equation}\label{eqn: GQ_Z is id times C}
    \hat{G}_{BT}\hat{Q}_Z = \id_B \otimes \hat{C}.
\end{equation}
This measurement scheme is depicted in Fig.~\ref{fig: virtual POVM}.

\begin{figure}
     \centering
     \includegraphics[width=0.75\linewidth]{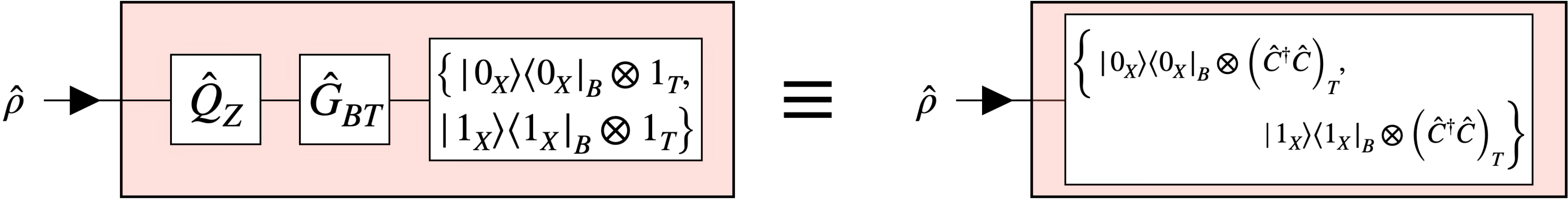}
     \caption{
     Bob's scheme for a successful measurement in the virtual scheme. Bob is allowed to perform any virtual operation on the incoming states in order to maximise his ability to guess Alice's virtual $X$ outcome. Here, this corresponds to apply the filter $\hat{Q}_Z$ to make sure the total number of detections in the $Z$ basis for the virtual and actual protocols are the same, and then an additional filter $\hat{G}_{BT}$, followed by projective measurements.
     A failed measurement in the virtual protocol is equivalently described by the POVM element $\id - \hat{M}_{BT}^{0_X,virt} - \hat{M}_{BT}^{1_X,virt}$ {or, equivalently,} the total filter $\hat{G}_{BT}\hat{Q}_{Z}$ not succeeding.}
     \label{fig: virtual POVM}
 \end{figure}

At round $l$, {if the virtual filtering is successful}, the joint probability of Alice sending the virtual state $\ket{\phi^{\pt{c}}}_B$ for $c\in\pg{1,2}$ and Bob obtaining $s_X^v$ (in the virtual scheme) for rounds where he actually measured in the $Z$ basis is given {by
\begin{align} \nonumber
    P_{s_X,c|O^{\pt{l-1}}}^{virt}
    &=\Tr\pq{\pt{\dyad{c}_A\otimes\hat{M}^{s_X,virt}_{BT} }\hat{\rho}_{ABT}^{l|O^{\pt{l-1}}} } \\ \nonumber
    &=\Tr\pq{\pt{\dyad{c}_A\otimes \dyad{s_X}_B\otimes \hat{C}^{\dag}\hat{C}}\hat{\rho}_{ABT}^{l|O^{\pt{l-1}}}} \\ \nonumber
    &=p_c\Tr\pq{\hat{D}_{s_X|O^{\pt{l-1}}}^{virt}\hat{P}\pt{\ket{\phi^{\pt{c}}}_B}}
    \\
    &=p_c\sum_{\omega \in \Omega} {W}^{\pt{c}}_{\omega} q_{s_X,\omega|O^{\pt{l-1}}}^{virt},
    \label{eqn: final for P virt virt}
\end{align}}
where we have defined
\begin{equation}\label{eqn: def of Dvirt}
    \hat{D}_{s_X|O^{\pt{l-1}}}^{virt} := \frac{1}{p^{\pt{l}}}\sum_{i,k,k'}\sum_{{\vec{x}_{<l}},{\vec{x}_{>l}}} \mel{\giklp}{\opxprod{\hat{C}}}{\gikl}_T \hat{A}^{\pt{\vec{x}_{<l},\vec{x}_{>l}}\dag}_{i,k';B|O^{\pt{l-1}}}\dyad{s_X}_B\hat{A}^{\pt{\vec{x}_{<l},\vec{x}_{>l}}}_{i,k;B|O^{\pt{l-1}}},
\end{equation}
and 
\begin{equation}\label{eqn: first definition of q_virt}
    q_{s_X,\omega|O^{\pt{l-1}}}^{virt} = \frac{1}{2} \Tr\pq{\hat{D}_{s_X|O^{\pt{l-1}}}^{virt}\hat{\sigma}_{\omega}}.
\end{equation}

As shown in Appendix~\ref{app: deriving nice form of q virt}, once we consider the sum over all rounds we can rewrite these terms in a more convenient form given {by
\begin{equation}\label{eqn: cute form of q_virt}
    \sum_{l=1}^N P_{s_X,c|O^{\pt{l-1}}}^{virt} 
    = p_c \sum_{\omega \in \Omega} {W}^{\pt{c}}_{\omega}
    \sum_{l=1}^N q_{s_X,\omega|O^{\pt{l-1}}}^{virt}
    = \frac{p_c}{2}\sum_{\omega \in \Omega} {W}^{\pt{c}}_{\omega} \Tr\pq{\hat{\rho}_E \pt{\hat{T}_{s_X,\omega} \otimes \opxprod{\hat{C}}}},
\end{equation}}
where $\hat{\rho}_E$ is a quantum state under full control of Eve that encodes all the information of her attack {(i.e., the terms $\hat{A}^{\pt{\vec{x}_{<l},\vec{x}_{>l}}}_{i;BT|O^{\pt{l-1}}}$ in Eq.~\eqref{eqn: def o A terms}) and, as such, its size depends on the number of rounds $N$. On the other hand, $\hat{T}_{s_X,\omega}$ is a $4\times4$ matrix whose analytic form is reported in Eqs.~\eqref{eqn: T_0_id}-\eqref{eqn: summary T_s_omega}.} In Sec.~\ref{sec: bound ep epprime} we show how these {latter quantities} can be bounded {from the} knowledge of the generalised efficiency operators $\opxprod{\hat{F}_s}$ and the terms $\pgt{q_{s_X,\omega}}_{s,\omega}$ learned from the LT approach.

For application in the next section and for better clarity, let us define from Eq.~\eqref{eqn: final for P virt virt} 
\begin{equation}\label{eqn: virtual yields}
    P_{s_X^v,0_X^v}^{\pt{l}} := P_{s_X,c=1|O^{\pt{l-1}}}^{virt} ,\quad P_{s_X^v,1_X^v}^{\pt{l}} := P_{s_X,c=2|O^{\pt{l-1}}}^{virt},
\end{equation}
such that $P_{s_X^v,a_X^v}^{\pt{l}}$ denotes the joint probability at round $l$ of Bob and Alice obtaining an outcome $s$ and $a$ in the virtual measurement scheme, respectively.

{\subsection{{Computing the secret key rate}}\label{sec: Computing the secret key rate}

Formally, in the asymptotic limit, the probability of being in sifted key round and the probability of having a bit error in $Z$ basis are respectively given by
\begin{equation}
    p_{Z}^{sift} = \lim_{N\to\infty} \frac{1}{N}\sum_{a,s=0}^1 N_{s_Z,a_Z}, \quad \quad p_{Z}^{err} := \lim_{N\to\infty} \frac{1}{N}\sum_{a\neq s} N_{s_Z,a_Z}.
\end{equation}
Note that both quantities are directly estimated by Alice in the protocol (see Sec.~\ref{sec: protocol}), and that the bit error rate can be directly computed from their ratio as indicated in Eq.~\eqref{eqn: e_b main thm}.}

As discussed in Appendix~\ref{sec: receiver}, the rounds with a detection in the $Z$ basis are the ones for which the filter $\hat{Q}_Z$ in Eq.~\eqref{eqn: Z filter} is successful. For these rounds, in the virtual protocol Bob applies the virtual filter $\hat{G}_{BT}$ {provided} in Eq.~\eqref{eqn: virtual filter introduction}. {The average of the probabilities of Bob's virtual $X$ measurement being successful (which are conditional on Alice and Bob's measurement outcomes in the previous rounds of the virtual protocol and include the probability of being in a sifted key round)} is given by
\begin{equation}\label{eqn: N_x_virt def}
    p_{X}^{virt} := \lim_{N\to\infty}\frac{1}{N}\sum_{a,s=0}^1 N_{s_X^v,a_X^v},
\end{equation}
where $N_{s_X^v,a_X^v}$ indicates the number of occurrences in which Bob obtains a virtual $X$ measurement outcome $s$ and Alice prepares the state $\ket{\phi_{a_X}^{virt}}_B$. 
Crucially, this is an average probability because Bob's virtual $X$ measurement consist of filtering operations, whose success probability is, in all generality, different at every round, since it depends on the specific states that the filter acts on.

With similar reasoning, the average probability of having a bit error in Bob's virtual $X$ measurement is given by
\begin{equation}\label{eqn: N_err_x_virt def}
    p_{X^{virt}}^{err} := \lim_{N\to\infty}\frac{1}{N}\sum_{a\neq s} N_{s_X^v,a_X^v}.
\end{equation}

{We note that the terms on the right-hand side of Eqs.~\eqref{eqn: N_x_virt def}-\eqref{eqn: N_err_x_virt def} can be estimated as the sum over the round index $l$ of the quantities $P_{s_X^v,a_X^v}^{\pt{l}}$ in Eq.~\eqref{eqn: virtual yields} for $N\to\infty$, thanks to Kato's inequality \cite{katoConcentrationInequality2020}.} As a result, we can define in the asymptotic regime the fraction of sifted key rounds in which {Bob's virtual $X$ measurement is successful as
\begin{equation}\label{eqn: ps introduction}
    r_{X}^{virt} := \frac{p_{X}^{virt}}
    {p_{Z}^{sift}} = \frac{1}
    {p_{Z}^{sift}}\lim_{N\to\infty}\frac{1}{N}
    \pq{\sum_{l=1}^N\pt{P_{0_X^v,0_X^v}^{\pt{l}} + P_{1_X^v,0_X^v}^{\pt{l}} + P_{0_X^v,1_X^v}^{\pt{l}} + P_{1_X^v,1_X^v}^{\pt{l}}}}.
\end{equation}}

Finally, we compute the phase-error rate $e_p$, which is the probability of Bob making an error in the guess of Alice's virtual $X$ outcome for the sifted rounds.  Following the same argument {of Eq.~\eqref{eqn: ps introduction}, in the asymptotic limit we have that
\begin{equation}\label{eqn: ep def}
    e_p := \frac{p_{X^{virt}}^{err}}{p_{X}^{virt}} = \frac{\lim_{N\to\infty}\frac{1}{N}\sum_{l=1}^N\pt{P_{0_X^v,1_X^v}^{\pt{l}} + P_{1_X^v,0_X^v}^{\pt{l}}}}{\lim_{N\to\infty}\frac{1}{N}\sum_{l=1}^N\pt{P_{0_X^v,0_X^v}^{\pt{l}} + P_{1_X^v,0_X^v}^{\pt{l}} + P_{0_X^v,1_X^v}^{\pt{l}} + P_{1_X^v,1_X^v}^{\pt{l}}}}.
\end{equation}}

{According to Koashi's security proof \cite{koashiSimpleSecurity2009}, these quantities allow Alice and Bob to compute how many sifted bits they must sacrifice for the sake of privacy amplification. In detail, to remove any residual information shared with Eve, {Alice and Bob must perform hashing on their sifted keys} for all the rounds in which Bob's virtual filter $\hat{G}_{BT}$ was not successful {(note that the average probability of the virtual filter failing is given by $p_{Z}^{sift}-p_{X}^{virt}$), plus additional rounds to counteract the erroneous estimates of Bob. Over $N$ rounds of the protocol, this last term is given by $N p_{X}^{virt} h_2\pt{e_p}$, where $h_2\pt{x}$ denotes the binary entropy, and the total number of hashing rounds required for privacy amplification is given by
\begin{equation}
    M_{PA} = Np_{Z}^{sift}\pt{1- r_{X}^{virt}} +
    Np_{Z}^{sift} r_{X}^{virt} h_2\pt{e_p}.
\end{equation}}
Moreover, Alice and Bob must sacrifice additional
\begin{equation}
    M_{EC} = f Np_{Z}^{sift} h_2\pt{e_b}
\end{equation}
sifted key bits to perform error correction (EC), with an EC code of efficiency $f$. Putting all considerations together, the {asymptotic} secret key rate per pulse sent can be retrieved to {be
\begin{align}
    R &\ge \lim_{N\to\infty}\frac{1}{N}\pt{ Np_{Z}^{sift} - M_{PA} - M_{EC} } \\
    &= {p_{Z}^{sift}} \pq{r_{X}^{virt}\pt{1-h_2\pt{e_p}} - fh_2\pt{e_b}}.
\end{align}}
By considering a lower bound on $r_{X}^{virt}$ and an upper bound on $e_p$, Eq.~\eqref{eqn: main theorem SKR} follows naturally.}

Crucially, the terms $P_{s_X^v,a_X^v}^{\pt{l}}$ in Eqs.~\eqref{eqn: ps introduction}-\eqref{eqn: ep def} depend on the choice of the operator $\hat{C}$ in the definition of the virtual filter. 
As the optimal choice of $\hat{C}$ might be subject to the specifics of the channel, in this work we consider for simplicity the operator $\hat{C}$ originally proposed in Ref.~\cite{fungSecurityProof2009}, which has been {obtained} optimising the noiseless case (that is, without {SPFs and attacks} on the $B$ system).
In the following section we show how this choice for $\hat{C}$ allows to directly bound the phase-error rate given the experimental data and a characterisation of the devices efficiency operators.

\subsection{Bounding the phase-error rate}\label{sec: bound ep epprime}
As illustrated in Fig.~\ref{fig: workflow}, to relate the phase-error rate to the statistics of the actual $X$ {basis} measurements we must find lower and upper bounds on the operator $\opxprod{\hat{C}}$ in the form given by Eq.~\eqref{eqn: summary op bound for CdagC}. One viable option to do so is to {employ} semidefinite programming, as discussed in Sec.~\ref{sec: main theorem}.
Alternatively, here we compute the analytical bounds provided by Eq.~\eqref{eqn: AB bounds for lambda main thm}.

From Eqs.~\eqref{eqn: main theorem diagonalization of FFFF}-\eqref{eqn: def C1}, we define $D_{min}$ ($D_{max}$) as the minimum (maximum) eigenvalue of $\hat{D}$. It follows that the minimum (maximum) eigenvalue of $\opxprod{\hat{C}_1}$ is
$\eta_{min} := \min\pt{D_{max}^{-1},1}$ ($\eta_{max} := \min\pt{D_{min}^{-1},1}$). Therefore, we have that
\begin{equation}
    \eta_{min}\id + \hat{C}' = \opxprod{\hat{C}_1} = \eta_{max}\id - \hat{C}'', 
\end{equation}
where $\hat{C}'$ and $\hat{C}''$ are positive semi-definite diagonal matrices. Therefore
\begin{align} 
    \opxprod{\hat{C}} &= \hat{F}_0^{\dag} \hat{U} \opxprod{\hat{C}_1} \hat{U}^{\dag} \hat{F}_0 
    \nonumber \\
    &= \eta_{min} \opxprod{\hat{F}_0} + \hat{F}_0^{\dag} \hat{U} \hat{C}' \hat{U}^{\dag} \hat{F}_0 
    \nonumber \\
    &\ge \eta_{min} \opxprod{\hat{F}_0} ,
    \label{eqn: lower bound for CC}
\end{align}
where the inequality holds because the term $\hat{F}_0^{\dag} \hat{U} \hat{C}' \hat{U}^{\dag} \hat{F}_0$ is positive semi-definite. Similarly, one finds
\begin{equation}\label{eqn: bound C with F0 max}
    \opxprod{\hat{C}} \le \eta_{max} \opxprod{\hat{F}_0}.
\end{equation}
Moreover, from Eq.~\eqref{eqn: main theorem diagonalization of FFFF} we have that
\begin{equation}
    \opxprod{\hat{F}_1} = 
    \hat{F}_0^{\dag}\hat{U}\hat{D}^{-1}\hat{U}^{\dag}\hat{F}_0, 
\end{equation}
which is in the same form as the first line of Eq.~\eqref{eqn: lower bound for CC}, with the substitutions $\opxprod{\hat{C}}\to\opxprod{\hat{F}_1}$ and $\opxprod{\hat{C}_1} \to \hat{D}^{-1}$. By applying the same reasoning, we obtain
\begin{equation}\label{eqn: bound F1 with F0}
    \frac{1}{D_{max}} \opxprod{\hat{F}_0} \le \opxprod{\hat{F}_1} \le \frac{1}{D_{min}} \opxprod{\hat{F}_0},
\end{equation}
which can be inverted to find bounds on $\opxprod{\hat{F}_0}$ in terms of $\opxprod{\hat{F}_1}$. 
The combination of the results of Eqs.~\eqref{eqn: lower bound for CC}-\eqref{eqn: bound C with F0 max} and~\eqref{eqn: bound F1 with F0} yields the analytical bounds in Eq.~\eqref{eqn: AB bounds for lambda main thm}.

Now, bounds in Eq.~\eqref{eqn: summary op bound for CdagC} are defined for the operator $\opxprod{\hat{C}} \in \mathcal{H}_T$ acting on the $T$ system. Nevertheless, it is straightforward to generalise them to operators in the form
\begin{equation}
    \hat{\Pi} \otimes \opxprod{\hat{C}} \in \mathcal{H}_4 \otimes \mathcal{H}_T,
\end{equation}
where $\hat{\Pi}$ is a $4\times4$ projector. As expectation values for projective measurements are bounded in $\pq{0,1}$, one finds
\begin{equation}\label{eqn: summary op bound for pi otimes CdagC}
    \lambda^{-}_s \pt{\hat{\Pi} \otimes \opxprod{\hat{F}_s}} \le
     \pt{\hat{\Pi} \otimes \opxprod{\hat{C}}} \le 
    \lambda^{+}_s  \pt{\hat{\Pi} \otimes \opxprod{\hat{F}_s}}.
\end{equation}

This is particularly relevant since the terms
$\hat{T}_{s_X,\omega}$ in Eq.~\eqref{eqn: cute form of q_virt} are a linear combination of projectors, as we show in Appendix~\ref{app: deriving nice form of q virt}. By paying attention to the sign these projectors come with in the definition of $\hat{T}_{s_X,\omega}$, one can conveniently apply upper and lower bounds. For example, consider the operator 
\begin{equation}
    \hat{T}_{0_X,{X}} = {\Xtilde{++} - \Xtilde{-+}},
\end{equation}
where ${\hat X}_{++}$ and ${\hat X}_{-+}$ are projectors (see Eq.~\eqref{eqn: definition tilde projectors}). It is possible to upper bound $\hat{T}_{0_X,{X}}$
by upper bounding its positive component and lower bounding its negative component as
\begin{align}
    \hat{T}_{0_X,\hat{X}} \otimes \opxprod{\hat{C}} &= \Xtilde{++}\otimes \opxprod{\hat{C}} - \Xtilde{-+} \otimes \opxprod{\hat{C}} \nonumber \\
    &\le \lambda^{+}_0 \pt{\Xtilde{++} \otimes \opxprod{\hat{F}_0}} - \lambda^{-}_0\pt{\Xtilde{-+} \otimes \opxprod{\hat{F}_0}}.
    \label{eqn: bounds on T_0x_X}
\end{align}
From Eq.~\eqref{eqn: bounds on T_0x_X}, one directly finds
\begin{align}
    \sum_{l=1}^N q_{0_X,X|O^{\pt{l-1}}}^{virt} &= \frac{1}{2}\Tr\pq{\hat{\rho}_E\pt{\hat{T}_{0_X,X} \otimes \opxprod{\hat{C}}}} \nonumber
    \\
    &\le \frac{\lambda_0^+}{2} \Tr\pq{\hat{\rho}_E\pt{\Xtilde{++} \otimes \opxprod{\hat{F}_0}}}  - \frac{\lambda_0^-}{2} \Tr\pq{\hat{\rho}_E\pt{\Xtilde{-+} \otimes \opxprod{\hat{F}_0}}} \nonumber
    \\
    &= \frac{\lambda_0^+}{2} \Tr\pq{\hat{\rho}_E\pt{\frac{\hat{T}_{0_X,\nohatid}+\hat{T}_{0_X,X}}{2} \otimes \opxprod{\hat{F}_0}}} \nonumber \\ 
    &\quad \quad - \frac{\lambda_0^-}{2} \Tr\pq{\hat{\rho}_E\pt{\frac{\hat{T}_{0_X,\nohatid}-\hat{T}_{0_X,X}}{2} \otimes \opxprod{\hat{F}_0}}} \nonumber
    \\
    &= \frac{\lambda_0^+ + \lambda_0^-}{4}\Tr\pq{\hat{\rho}_E\pt{\hat{T}_{0_X,X} \otimes \opxprod{\hat{F}_0}}} \nonumber \\ 
    &\quad \quad + \frac{\lambda_0^+ - \lambda_0^-}{4}\Tr\pq{\hat{\rho}_E\pt{\hat{T}_{0_X,\nohatid} \otimes \opxprod{\hat{F}_0}}}
    \nonumber
    \\
    &= \frac{\lambda_0^+ + \lambda_0^-}{2}\sum_{l=1}^N q_{0_X,X|O^{\pt{l-1}}} + \frac{\lambda_0^+ - \lambda_0^-}{2}\sum_{l=1}^N q_{0_X,\nohatid|O^{\pt{l-1}}}.
\end{align}

Reasoning in a similar way, {we obtain} that for $s\in\pgt{0,1}$ it holds
\begin{equation}\label{eqn: bounds on q_s_id}
    \lambda_s^- \sum_{l=1}^N q_{s_X,\nohatid|O^{\pt{l-1}}} \le \sum_{l=1}^N q_{s_X,\nohatid|O^{\pt{l-1}}}^{virt} \le \lambda_s^+ \sum_{l=1}^N q_{s_X,\nohatid|O^{\pt{l-1}}},
\end{equation}
and 
\begin{align} \nonumber 
    \frac{\lambda_s^+ + \lambda_s^-}{2} \sum_{l=1}^N q_{s_X,X|O^{\pt{l-1}}} - \frac{\lambda_s^+ - \lambda_s^-}{2} &\sum_{l=1}^N q_{s_X,\nohatid|O^{\pt{l-1}}}
    \\ \nonumber
    \le
    \sum_{l=1}^N &q_{s_X,X|O^{\pt{l-1}}}^{virt} 
    \le
    \\
    \frac{\lambda_s^+ + \lambda_s^-}{2} &\sum_{l=1}^N q_{s_X,X|O^{\pt{l-1}}} + \frac{\lambda_s^+ - \lambda_s^-}{2} \sum_{l=1}^N q_{s_X,\nohatid|O^{\pt{l-1}}}.\label{eqn: bounds on q_s_X}
\end{align}

Note that for detectors with equal efficiencies, $\opxprod{\hat{F}_0} = \opxprod{\hat{F}_1} \implies \hat{D} = \id$, hence $\lambda_s^{\pm} = 1$ and the virtual terms $q$ are the same as in the actual $X$ measurement. As a result, we are back to the scenario of the original LT protocol and the phase-error rate is zero, as expected \cite{tamakiLosstolerantQuantum2014}. Moreover, in the general case the phase-error rate given by Eq.~\eqref{eqn: ep def} can be directly upper bounded in a LT fashion, given the results of Eqs.~\eqref{eqn: bounds on q_s_id}-\eqref{eqn: bounds on q_s_X}. In fact, recalling the values of $p_1$ and $p_2$ from Table~\ref{tab: virtual protocol recap}, we have that
\begin{align}\nonumber
    \sum_{l=1}^N\pt{P_{0_X^v,1_X^v}^{\pt{l}} + P_{1_X^v,0_X^v}^{\pt{l}}} &= p_{1_X^{virt}}p_{Z_A}p_{Z_B}\sum_{l=1}^N\pt{q_{0_X,\id|O^{\pt{l-1}}}^{virt} - q_{0_X,X|O^{\pt{l-1}}}^{virt}} \\\nonumber
    & \quad + p_{0_X^{virt}}p_{Z_A}p_{Z_B}\sum_{l=1}^N\pt{q_{1_X,\id|O^{\pt{l-1}}}^{virt} + q_{1_X,X|O^{\pt{l-1}}}^{virt}} \\\nonumber
    &\le p_{1_X^{virt}}p_{Z_A}p_{Z_B}\sum_{l=1}^N\pt{\frac{3\lambda_0^+ - \lambda_0^-}{2}q_{0_X,\id|O^{\pt{l-1}}} - \frac{\lambda_0^+ + \lambda_0^-}{2}   q_{0_X,X|O^{\pt{l-1}}}} \\
    & \quad + p_{0_X^{virt}}p_{Z_A}p_{Z_B}\sum_{l=1}^N\pt{\frac{3\lambda_1^+ - \lambda_1^-}{2}q_{1_X,\id|O^{\pt{l-1}}} + \frac{\lambda_1^+ + \lambda_1^-}{2}   q_{1_X,X|O^{\pt{l-1}}}}.
    \label{eqn: upper bound on ep numerator}
\end{align}

Similarly, we have that
\begin{align}\nonumber
    \sum_{l=1}^N&\pt{P_{0_X^v,1_X^v}^{\pt{l}} + P_{1_X^v,1_X^v}^{\pt{l}} + P_{0_X^v,0_X^v}^{\pt{l}} + P_{1_X^v,0_X^v}^{\pt{l}}} 
    \\\nonumber
    &= p_{1_X^{virt}}p_{Z_A}p_{Z_B}\sum_{l=1}^N\pt{q_{0_X,\id|O^{\pt{l-1}}}^{virt} - q_{0_X,X|O^{\pt{l-1}}}^{virt} + q_{1_X,\id|O^{\pt{l-1}}}^{virt} - q_{1_X,X|O^{\pt{l-1}}}^{virt}} \\\nonumber
    & \quad + p_{0_X^{virt}}p_{Z_A}p_{Z_B}\sum_{l=1}^N\pt{q_{0_X,\id|O^{\pt{l-1}}}^{virt} + q_{0_X,X|O^{\pt{l-1}}}^{virt} + q_{1_X,\id|O^{\pt{l-1}}}^{virt} + q_{1_X,X|O^{\pt{l-1}}}^{virt}} \\\nonumber
    &\ge 
    p_{1_X^{virt}}p_{Z_A}p_{Z_B}\sum_{l=1}^N\pt{
    -\frac{\lambda_0^+ - 3\lambda_0^-}{2} q_{0_X,\id|O^{\pt{l-1}}} - \frac{\lambda_0^+ +\lambda_0^-}{2}q_{0_X,X|O^{\pt{l-1}}}}
    \\\nonumber
    & \quad +
    p_{1_X^{virt}}p_{Z_A}p_{Z_B}\sum_{l=1}^N\pt{
      -\frac{\lambda_1^+ - 3\lambda_1^-}{2}q_{1_X,\id|O^{\pt{l-1}}} - \frac{\lambda_1^+ +\lambda_1^-}{2}q_{1_X,X|O^{\pt{l-1}}}}
    \\\nonumber
    & \quad + p_{0_X^{virt}}p_{Z_A}p_{Z_B}\sum_{l=1}^N\pt{
    -\frac{\lambda_0^+ - 3\lambda_0^-}{2} q_{0_X,\id|O^{\pt{l-1}}} + \frac{\lambda_0^+ +\lambda_0^-}{2}q_{0_X,X|O^{\pt{l-1}}}},
    \\\nonumber
    & \quad + p_{0_X^{virt}}p_{Z_A}p_{Z_B}\sum_{l=1}^N\pt{ -\frac{\lambda_1^+ - 3\lambda_1^-}{2}q_{1_X,\id|O^{\pt{l-1}}} + \frac{\lambda_1^+ +\lambda_1^-}{2}q_{1_X,X|O^{\pt{l-1}}}},
    \\\nonumber
    &= p_{Z_A}p_{Z_B}\sum_{l=1}^N\left[-\frac{\lambda_0^+ - 3\lambda_0^-}{2} q_{0_X,\id|O^{\pt{l-1}}} -\frac{\lambda_1^+ - 3\lambda_1^-}{2}q_{1_X,\id|O^{\pt{l-1}}} + \right.
    \\
    &\quad \left.+\braket{\phi_{0_Z}}{\phi_{1_Z}}\pt{\frac{\lambda_0^+ +\lambda_0^-}{2}q_{0_X,X|O^{\pt{l-1}}} + \frac{\lambda_1^+ +\lambda_1^-}{2}q_{1_X,X|O^{\pt{l-1}}}}\right].
    \label{eqn: lower bound on ep denom}
\end{align}
Note that in the last passage we retrieved the explicit form of the probability of sending the virtual states in Eq.~\eqref{eqn: prob of sending 0x virt}.

{By dividing by $N$ and taking the limit $N\to\infty$ of Eqs.~\eqref{eqn: upper bound on ep numerator}-\eqref{eqn: lower bound on ep denom}, and combining with the results of Eq.~\eqref{eqn: qtilda def} and Eqs.~\eqref{eqn: N_x_virt def}-\eqref{eqn: N_err_x_virt def}, we can find the bounds on {$p_X^{virt}$} and $p_{X^{virt}}^{err}$ reported in Eqs.~\eqref{eqn: main theorem N_x_virt}-\eqref{eqn: main thm bound e_p}. From here, the bounds $r^{virt,L}_X$ and  $e_p^{U}$ in Eqs.~\eqref{eqn: main thm bound p_s}-\eqref{eqn: main thm bound e_p} follow.}
Importantly, all quantities involved are known and include {SPFs} ($\braket{\phi_{0_Z}}{\phi_{1_Z}} \neq 0$), {DEM} (in the values of the parameters $\lambda_s^{\pm}$ for $s\in\pg{0,1}$) and the action of the channel (in the terms $\pg{q_{s_X,\omega}}_{\omega}$).

\section{Derivation of Eq.~\eqref{eqn: cute form of q_virt}}\label{app: deriving nice form of q virt}
Consider the following expansions of operators on the system $B$ in terms of the Pauli basis. We have
\begin{gather}
    \dyad{s_X}_B = \frac{1}{2}\pt{\id + \pt{-1}^s\hat{X}}, 
    \\
    \hat{A}^{\pt{\vec{x}_{<l},\vec{x}_{>l}}}_{i,k;B|O_{l-1}} = a_{0}^{\pt{l,i,k}} \id + a_{1}^{\pt{l,i,k}} \hat{X} + i a_{2}^{\pt{l,i,k}} \hat{Y} + a_{3}^{\pt{l,i,k}} \hat{Z} ,
    \\
    \hat{A}^{\pt{\vec{x}_{<l},\vec{x}_{>l}}\dag}_{i,k';B|O_{l-1}} = a_{0}^{\pt{l,i,k'}*} \id + a_{1}^{\pt{l,i,k'}*} \hat{X} -i a_{2}^{\pt{l,i,k'}*} \hat{Y} + a_{3}^{\pt{l,i,k'}*} \hat{Z}.
\end{gather} 
In the following, let $a_{t}^{\pt{l,i,k}} \equiv a_t$ and $a_{t}^{\pt{l,i,k'}*} \equiv {a'}_{t}^{*}$ with $t\in\pg{0,1,2,3}$ for ease of {notation}. 

By exploiting the properties of the trace of the product of Pauli matrices, with the notation introduced in Eq.~\eqref{eqn: def sigma op as pauli matrices} we notice that
\begin{equation}\label{eqn: factorisation of q, B subsys, X component}
\Tr\pq{\hat{A}^{\pt{\vec{x}_{<l},\vec{x}_{>l}}\dag}_{i,k';B|O_{l-1}}\dyad{s_X}_B\hat{A}^{\pt{\vec{x}_{<l},\vec{x}_{>l}}}_{i,k;B|O_{l-1}} \hat{\sigma}_{\omega}} = E_{\nohatid,{\omega}}^{\pt{l,i,k,k'}} + \pt{-1}^s E_{X,{\omega}}^{\pt{l,i,k,k'}},
\end{equation}
where $\omega\in\pg{\nohatid,X,Z}$ and
\begin{equation}
\begin{gathered}
     E_{\nohatid,\nohatid}^{\pt{l,i,k,k'}} =  {\apa{0}{0} + \apa{1}{1} + \apa{2}{2} + \apa{3}{3}},  \\
    E_{\nohatid,X}^{\pt{l,i,k,k'}} =  {\apa{1}{0} + \apa{0}{1} + \apa{2}{3} + \apa{3}{2}} , \\
    E_{\nohatid,Z}^{\pt{l,i,k,k'}} =  {\apa{3}{0} + \apa{0}{3} - \apa{1}{2} -  \apa{2}{1}} ,\\
    E_{X,\nohatid}^{\pt{l,i,k,k'}} = {\apa{1}{0} + \apa{0}{1} - \apa{3}{2} - \apa{2}{3}} ,\\
    E_{X,X}^{\pt{l,i,k,k'}} = {\apa{0}{0} + \apa{1}{1} - \apa{2}{2} - \apa{3}{3}},  \\
    E_{X,Z}^{\pt{l,i,k,k'}} =  {\apa{3}{1} + \apa{1}{3} - \apa{0}{2} - \apa{2}{0}}.
\end{gathered}
\end{equation}

Now, let
\begin{equation}\label{eqn: definition tilde projectors}
\begin{aligned}[c]
    \Ztilde{00} &:= \hat{P}\pt{\pq{1,0,0,1}^{\dag}}/2, \\
    \Ztilde{10} &:= \hat{P}\pt{\pq{0,1,1,0}^{\dag}}/2, \\
       \Xtilde{++} &:= \hat{P}\pt{\pq{1,1,0,0}^{\dag}}/2, \\ 
       \Xtilde{-+} &:= \hat{P}\pt{\pq{0,0,-1,1}^{\dag}}/2, \\
       \Ytilde{++} &:= \hat{P}\pt{\pq{1,0,1,0}}/2, \\
       \Ytilde{-+} &:= \hat{P}\pt{\pq{0,-1,0,1}}/2, 
\end{aligned}
\quad \quad \quad \quad \quad
\begin{aligned}[c]
       \Ztilde{01} &:= \hat{P}\pt{\pq{0,1,-1,0}^{\dag}}/2, \\
       \Ztilde{11} &:= \hat{P}\pt{\pq{1,0,0,-1}^{\dag}}/2, \\
       \Xtilde{+-} &:= \hat{P}\pt{\pq{0,0,1,1}^{\dag}}/2, \\
       \Xtilde{--} &:= \hat{P}\pt{\pq{1,-1,0,0}^{\dag}}/2, \\
       \Ytilde{+-} &:= \hat{P}\pt{\pq{0,1,0,1}}/2, \\
       \Ytilde{--} &:= \hat{P}\pt{\pq{1,0,-1,0}}/2.
\end{aligned}
\end{equation}

We notice that
\begin{align}
E_{\nohatid,\nohatid}^{\pt{l,i,k,k'}} + E_{X,\nohatid}^{\pt{l,i,k,k'}} &= 
    \apa{0}{0} + \apa{1}{1} + \apa{1}{0} + \apa{0}{1} + \apa{2}{2} + \apa{3}{3} -\apa{3}{2} - \apa{2}{3}
    \nonumber \\
    &= \pt{{a'_0}^{*} + {a'_1}^{*}}\pt{a_0 + a_1} + \pt{-{a'_2}^{*} + {a'_3}^{*}}\pt{-a_2 + a_3}  \nonumber \\
    &= {B^{\pt{l,i}}_{k'}}^{\dag}\pq{\Xtilde{++} + \Xtilde{-+}}B^{\pt{l,i}}_k \nonumber \\
    &= {B^{\pt{l,i}}_{k'}}^{\dag}\hat{T}_{0_X,\id} B^{\pt{l,i}}_k,
\end{align}
where we defined 
\begin{equation}
    B^{\pt{l,i}}_k := \mqty(a_0^{\pt{l,i,k}}&a_1^{\pt{l,i,k}}&a_2^{\pt{l,i,k}}&a_3^{\pt{l,i,k}})^{T},
\end{equation}
and 
\begin{equation}\label{eqn: T_0_id}
    \hat{T}_{0_X,\nohatid} := {\Xtilde{++} + \Xtilde{-+}}.
\end{equation}

We can proceed similarly for all terms in the form of Eq.~\eqref{eqn: factorisation of q, B subsys, X component} to find 
\begin{equation}
    E_{\nohatid,\omega}^{\pt{l,i,k,k'}} + \pt{-1}^s E_{X,\omega}^{\pt{l,i,k,k'}} = {B^{\pt{l,i}}_{k'}}^{\dag} \hat{T}_{s_X,\omega}B^{\pt{l,i}}_k,
\end{equation}
with 
\begin{equation}\label{eqn: summary T_s_omega}
\begin{gathered}
\hat{T}_{1_X,\nohatid} = {\Xtilde{--} + \Xtilde{+-}} , \nonumber \\
\hat{T}_{0_X,{X}} = {\Xtilde{++} - \Xtilde{-+}} , \nonumber \\
\hat{T}_{1_X,{X}} = {- \Xtilde{--} + \Xtilde{+-}} , \nonumber \\
\hat{T}_{0_X,{Z}} = {\Ztilde{00} + \Ztilde{01} - \Ytilde{-+} - \Ytilde{++}} , \nonumber \\
\hat{T}_{1_X,{Z}} = {- \Ztilde{11} - \Ztilde{10} + \Ytilde{-+} + \Ytilde{++}}.
\end{gathered}
\end{equation}

Let now 
\begin{equation}
    \hat{B}^{\pt{l,i}} = \mqty(
    a_0^{\pt{l,i,0}} & a_0^{\pt{l,i,1}} & \ldots \\
    a_1^{\pt{l,i,0}} & a_1^{\pt{l,i,1}} & \ldots \\
    a_2^{\pt{l,i,0}} & a_2^{\pt{l,i,1}} & \ldots \\
    a_3^{\pt{l,i,0}} & a_3^{\pt{l,i,1}} & \ldots \\
    ) = \mqty( &  & \\ &  & \\
    B^{\pt{l,i}}_0 & B^{\pt{l,i}}_1 & \ldots \\
     &  & \\ &  & \\),
\end{equation}
and 
\begin{equation}
    \ket{\Gamma^{\pt{l,i}}} = \mqty(\ket{\gamma^{\pt{l,i,0}}}\\\ket{\gamma^{\pt{l,i,1}}}\\\vdots).
\end{equation}
One can easily check that
\begin{equation}
    {B^{\pt{l,i}}_{k'}}^{\dag} \hat{T}_{s_X,{\omega}} B^{\pt{l,i}}_k = \pt{{\hat{B}}^{\pt{l,i}\dag} \hat{T}_{s_X,{\omega}}\hat{B}^{\pt{l,i}}}_{k'k},
\end{equation}
and consequently, from Eq.~\eqref{eqn: first definition of q_virt} we have that
\begin{align}
    q_{s_X,\omega}^{virt,\pt{l}} &= \frac{1}{2} \sum_{i,k,k'} \mel{\giklp}{\opxprod{\hat{C}}}{\gikl}_T\Tr\pq{\hat{A}^{\pt{\vec{x}_{<l},\vec{x}_{>l}}\dag}_{i,k';B|O_{l-1}}\dyad{s_X}_B\hat{A}^{\pt{\vec{x}_{<l},\vec{x}_{>l}}}_{i,k;B|O_{l-1}} \hat{\sigma}_{\omega}} \nonumber \\
    &= \frac{1}{2} \sum_{i,k,k'} \mel{\giklp}{\opxprod{\hat{C}}}{\gikl}_T \pt{{\hat{B}}^{\pt{l,i}\dag} \hat{T}_{s_X,{\omega}}\hat{B}^{\pt{l,i}}}_{k'k} \nonumber \\
    &= \frac{1}{2} \sum_i \expval{\pq{{\hat{B}}^{\pt{l,i}\dag} \hat{T}_{s_X,{\omega}}\hat{B}^{\pt{l,i}} \otimes \opxprod{\hat{C}}}}{\Gamma^{\pt{l,i}}} \nonumber \\
    &= \frac{1}{2} \sum_i \expval{ \hat{T}_{s_X,{\omega}} \otimes \opxprod{\hat{C}}}{\Phi^{\pt{l,i}}},
    \label{eqn: q virt final tensor expval}
\end{align}
where we have defined (recall Eq.~\eqref{eqn: unnormalised gamma T})
\begin{equation}
    \ket{\Phi^{\pt{l,i}}} := \pt{\hat{B}^{\pt{l,i}} \otimes \id_T} \ket{\Gamma^{\pt{l,i}}} = \mqty(
    a_0^{\pt{l,i,0}}\hat{A}^{\pt{\vec{x}_{<l},\vec{x}_{>l}}}_{i,0;B|O_{l-1}} & a_0^{\pt{l,i,1}}\hat{A}^{\pt{\vec{x}_{<l},\vec{x}_{>l}}}_{i,1;B|O_{l-1}} & \ldots \\
    a_1^{\pt{l,i,0}}\hat{A}^{\pt{\vec{x}_{<l},\vec{x}_{>l}}}_{i,0;B|O_{l-1}} & a_1^{\pt{l,i,1}}\hat{A}^{\pt{\vec{x}_{<l},\vec{x}_{>l}}}_{i,1;B|O_{l-1}} & \ldots \\
    a_2^{\pt{l,i,0}}\hat{A}^{\pt{\vec{x}_{<l},\vec{x}_{>l}}}_{i,0;B|O_{l-1}} & a_2^{\pt{l,i,1}}\hat{A}^{\pt{\vec{x}_{<l},\vec{x}_{>l}}}_{i,1;B|O_{l-1}} & \ldots \\
    a_3^{\pt{l,i,0}}\hat{A}^{\pt{\vec{x}_{<l},\vec{x}_{>l}}}_{i,0;B|O_{l-1}} & a_3^{\pt{l,i,1}}\hat{A}^{\pt{\vec{x}_{<l},\vec{x}_{>l}}}_{i,1;B|O_{l-1}} & \ldots \\
    )\mqty(\ket{0}_T\\\vdots\\\ket{0}_T).
\end{equation}

By taking the sum over $l$ of Eq.~\eqref{eqn: q virt final tensor expval} and letting 
\begin{equation}
    \hat{\rho}_E := \sum_{l,i}\dyad{\Phi^{\pt{l,i}}},
\end{equation}
one {obtains} the result in Eq.~\eqref{eqn: cute form of q_virt}.

\bibliographystyle{apsrev4-2.bst}
\bibliography{MyLibrary}

\end{document}